\documentclass[lettersize,journal]{IEEEtran}
\usepackage{amsmath,amsfonts}
\usepackage{algorithmicx,algorithm}
\usepackage{array}
\usepackage[caption=false,font=normalsize,labelfont=sf,textfont=sf]{subfig}
\usepackage{textcomp}
\usepackage{stfloats}
\usepackage{url}
\usepackage{algpseudocode}
\usepackage{verbatim}
\usepackage{graphicx}
\usepackage{color}
\usepackage{amssymb}
\usepackage{bm}
\usepackage{epstopdf}
\usepackage{multirow}
\usepackage{cite}
\usepackage{fancyhdr}
\usepackage{booktabs}
\usepackage{hyperref}
\usepackage{float}
\newtheorem{lemma}{\textbf{Lemma}}

\def\BibTeX{{\rm B\kern-.05em{\sc i\kern-.025em b}\kern-.08em
    T\kern-.1667em\lower.7ex\hbox{E}\kern-.125emX}}
\usepackage{balance}
\begin{document}
\title{{Nest-DGIL: Nesterov-optimized Deep Geometric Incremental Learning for CS Image Reconstruction}}
\author{Xiaohong Fan, Yin Yang, Ke Chen, Yujie Feng, and Jianping Zhang
\thanks{This work was supported by the National Natural Science Foundation of China Project (11771369, 12071402, 12261131501), the Education Bureau of Hunan Province, P. R. China (22A0119), the National Key Research and Development Program of China (2020YFA0713503), the Natural Science Foundation of Hunan Province (2020JJ2027, 2020ZYT003, 2023GK2029), the Project of Scientific Research Fund of the Hunan Provincial Science and Technology Department (2022RC3022), the Postgraduate Scientific Research Innovation Project of Hunan Province (CX20210598) and Postgraduate Scientific Research Innovation Project of Xiangtan University (XDCX2021B097), P. R. China. (Corresponding author: Jianping Zhang).}
\thanks{X. Fan and Y. Feng are with the School of Mathematics and Computational Science, Xiangtan University, and Hunan Key Laboratory for Computation and Simulation in Science and Engineering, Xiangtan 411105, China (fanxiaohong@smail.xtu.edu.cn, fengyujie@smail.xtu.edu.cn).}
\thanks{Y. Yang is with the School of Mathematics and Computational Science, Xiangtan University, and Hunan National Applied Mathematics Center, Xiangtan 411105, China (yangyinxtu@xtu.edu.cn).}
\thanks{K. Chen is with the Centre for Mathematical ImagingTechniques, Department of Mathematical Sciences, The University of Liverpool, Liverpool,
and Department of Mathematics and Statistics, University of Strathclyde, Glasgow, U.K (k.chen@strath.ac.uk).}
\thanks{J. Zhang is with the School of Mathematics and Computational Science, Xiangtan University, and the Key Laboratory of Intelligent Computing \& Information Processing of the Ministry of Education, Xiangtan 411105, China. (jpzhang@xtu.edu.cn).}
}

\markboth{Journal of \LaTeX\ Class Files,~Vol.~18, No.~9, September~2020}%
{How to Use the IEEEtran \LaTeX \ Templates}

\maketitle

\begin{abstract}
Proximal gradient-based optimization is one of the most common strategies to solve inverse problem of images, and it is easy to implement. However, these techniques often generate heavy artifacts in image reconstruction. One of the most popular refinement methods is to fine-tune the regularization parameter to alleviate such artifacts, but it may not always be sufficient or applicable due to increased computational costs. In this work, we propose a deep geometric incremental learning framework based on the second Nesterov proximal gradient optimization. The proposed end-to-end network not only has the powerful learning ability for high-/low-frequency image features, but also can theoretically guarantee that geometric texture details will be reconstructed from preliminary linear reconstruction. Furthermore, it can avoid the risk of intermediate reconstruction results falling outside the geometric decomposition domains and achieve fast convergence. Our reconstruction framework is decomposed into four modules including general linear reconstruction, cascade geometric incremental restoration, Nesterov acceleration, and post-processing. In the image restoration step, a cascade geometric incremental learning module is designed to compensate for missing texture information from different geometric spectral decomposition domains. Inspired by the overlap-tile strategy, we also develop a post-processing module to remove the block effect in patch-wise-based natural image reconstruction. All parameters in the proposed model are learnable, an adaptive initialization technique of physical parameters is also employed to make model flexibility and ensure converging smoothly. We compare the reconstruction performance of the proposed method with existing state-of-the-art methods to demonstrate its superiority. Our source codes are available at \emph{https://github.com/fanxiaohong/Nest-DGIL}.
\end{abstract}


\section{Introduction}
\IEEEPARstart{I}{mage} reconstruction is one of the most basic problems in computer vision and is also a challenging problem in the medical imaging community. One of the most popular approaches is Compressed Sensing (CS) reconstruction, which refers to the process of reconstructing an image $\bm{x}\in \mathbb{R}^{N}$ from the imaging system defined by
\begin{equation}
\boldsymbol{\Phi} \bm{x}=\bm{y},
\label{eq0}
\end{equation}
where an observation $\bm{y}\in\mathbb{R}^{M}$ ($M=m_1m_2$) has been sampled significantly below Shannon-Nyquist rate \cite{Candes2008,Candes2006a,Candes2006,Donoho2006}, $\bm{x}$ and $\bm{y}$ are resized from $n_1\times n_2$ and $m_1\times m_2$ images respectively. It has been widely used in single-pixel cameras \cite{Duarte2008}, accelerating magnetic resonance imaging (MRI) \cite{Lustig2007}, sparse-view computational tomography (CT) \cite{Xiang2020}, high-speed videos \cite{Hitomi2011} and other fields. A new class of reconstruction methods for robust structured compressible signal recovery is proposed \cite{Baraniuk2010}. Therefore, CS reconstruction has become a powerful image compressing and reconstructing tool.

The task \eqref{eq0} of reconstructing an unknown $\bm{x}$ from under-sampled observation $\bm{y}$, is ill-posed due to the under-determination of the linear measurement matrix $\boldsymbol{\Phi}\in\mathbb{R}^{M\times N}$ ($M \ll N$). Therefore, the classical mathematical model of image reconstruction can be given by
\begin{equation}
\underset{\bm{x}\in\mathcal{X}}{\min } \left\{\mathcal{S}(\bm{x})+\lambda\mathcal{R}(\bm{x})\right\},
\label{eq1a}
\end{equation}
where $\mathcal{R}(\bm{x})$ is a regularizer with image geometric prior (\emph{e.g.} sparsity and smoothness etc.), $\lambda$ is a regularization parameter that balances data fidelity term $\mathcal{S}(\bm{x})=\frac{1}{2}\left\|\boldsymbol{\Phi} \bm{x}-\bm{y}\right\|_{2}^{2}$ and image prior $\mathcal{R}(\bm{x})$. The CS ratio is defined as $\frac{M}{N}$, $\mathcal{X}$ is a kind of image space.
$\mathcal{R}(\bm{x})$ can be carried out in image domain or traditional transform domain (such as gradient domain, wavelet domain, etc.) \cite{Daubechies2004,Dong2014,He2009,Qu2012,Ravishankar2011}, however it cannot adequately represent the complex textures of the image.

To solve \eqref{eq1a}, many iterative algorithms have been developed, such as the iterative shrinkage-thresholding algorithm (ISTA) \cite{Daubechies2004,Elad2006}, the fast iterative shrinkage-thresholding algorithm (FISTA) \cite{Beck2009}, the alternating direction multiplier method (ADMM) \cite{Boyd2010}, and so on. These methods are based on interpretable and predefined image geometric prior rather than learning directly; most of them have the advantages of theoretical analysis and strong convergence. However, they not only take hundreds of iterations to converge, but also have to face the difficulty of choosing optimal geometric prior $\mathcal{R}(\bm{x})$ and physical parameters, which may lead to nonoptimal image reconstruction results.

In fact, the optimal solution of \eqref{eq1a} can be given by
\begin{equation}
g(\bm{x}):=\boldsymbol{\Phi}^T\boldsymbol{\Phi}\bm{x}+\lambda\mathcal{R}^\prime(\bm{x})=\boldsymbol{\Phi}^T\bm{y}.
\label{eq2b}
\end{equation}
Especially if the operator $g(\cdot)$ is an one-to-one smooth mapping, thus the optimization algorithm is seeking a suitable non-linear reconstruction mapping $f(\cdot)$ to achieve image reconstruction, which is denoted by
 \[\bm{x}=f(\bm{y})=g^{-1}(\boldsymbol{\Phi}^T\bm{y}).\]
It is well known that high-quality reconstruction consists of three key factors, i.e. the geometric prior of $\bm{x}$ in image space $\mathcal{X}$, non-linear mapping $f$ and measurement matrix $\Phi$. In this work, thus a neural network can be learned to solve the optimization problem defined by
 \begin{equation}
\mathcal{E}(\mathcal{X})=\inf_{f} \sup_{\bm{x} \in \mathcal{X}}\|\bm{x}-f(\boldsymbol{\Phi} \bm{x})\|. 
\label{eq0b}
\end{equation}
The popular convolutional neural network (CNN) is suitable for learning the non-linear reconstruction mapping $f$ directly in training space $\mathcal{X}$ \cite{Kulkarni2016,Xie2018,Mardani2019,Mousavi2017,Yang2018,Zhong2022}. Recently, the regularization model architecture has been embedded into the deep convolutional neural network (DnCNN) for image denoising \cite{Zhang2017a}. With the observation that unrolled iterative methods have the form of a CNN (filtering followed by pointwise nonlinearity), an indirect inversion approximated by a CNN is proposed to solve normal-convolutional inverse problems \cite{Jin2017}, such network combines multi-resolution decomposition and residual learning to remove undesirable artifacts while preserving image geometries \cite{He2016}. An untrained image reconstruction framework called the Deep Decoder is proposed to generate natural images by using very few weight parameters\cite{Heckel2019}. CSformer is a hybrid end-to-end CS framework, which is composed of adaptive sampling and recovery, to explore the representation capacity of local and global features \cite{Ye2023}. In contrast to many classical model-based methods with sparsity theory, learning methods can dramatically reduce computational complexity and achieve impressive reconstruction performance. However, these existing deep learning-based methods are trained as a black box and are driven by massive training data, with limited theoretical insights from geometric characteristic domains.

There are many approaches in the literature for formulating and designing a CNN architecture in terms of interpretable components. Learned Iterative
Shrinkage Thresholding Algorithm (LISTA) is first
proposed to learn optimal sparse codes \cite{Gregor2010}. ISTA algorithm is embedded into deep network (ISTA-Net and ISTA-Net+), and the aim is to learn proximal mapping using nonlinear transforms \cite{Zhang2018}. FISTA-Net that consists of gradient descent, proximal mapping, and two-step update, is designed by mapping the FISTA algorithm into a deep network \cite{Xiang2020}. AMP-Net is established by unfolding the iterative denoising process of the well-known approximate message passing (AMP) algorithm rather than learning regularization terms \cite{Zhang2021}. These networks designing deep architectures have theoretical support and can alleviate the instability of pure data-driven methods; we also refer the readers to \cite{yangyan2016,Lempitsky2018,Adler2018,Aggarwal2019,Duan2019,Borgerding2017,Metzler2017,Yang2018a,Liu2020TCI,Zhang2020a,You2021,Chen2021,Fan2021,Alver2021,Hou2022,Yang2022,Zhang2023,Fan2023} for more details.

However, existing deep unfolding methods still have the drawback of an insufficiently theoretical relationship between optimization theory and network architecture design. ISTA-Net+ and FISTA-Net have neglected other regularization priories (just $\ell_{1}$ prior information), the nonlinear transform and its inverse transform are directly replaced by several convolution layers, which leads to a not straight reasonable explanation. Although AMP-Net iterates the denoising process rather than learning regularization terms, the CNN denoiser is not well analyzed and explained. Unfortunately, the two-step update, which is a well-designed linear combination of previous reconstructions in FISTA-Net, runs the risk of being outside the geometric domains and losing meaning \cite{Nesterov1988,Liu2021}.

To overcome above drawbacks of existing deep unfolding methods, inspired by the operator spectral decomposition, we reformulate a new geometric characteristic series dealing with the proximal-point update of the classical reconstruction, to develop
a deep Geometric Incremental Learning (GIL) framework based on the second Nesterov proximal gradient optimization in this work. The derivation from mathematical theory to our network design is natural and explainable. Our contributions can be summarized as follows.
\begin{itemize}
\item We propose a Nesterov-informed Deep Geometric Incremental Learning (Nest-DGIL) framework, which has the powerful learning ability for high/low frequency image features and can theoretically guarantee that more geometric texture details will be reconstructed from preliminary linear reconstruction. Such a network gives us a new perspective on how to design an explainable architecture.
\item A cascade GIL module, which is inspired by a geometric spectral decomposition of the nonlinear inverse operator and combines with a multi-regularisers truncation restoration, is designed to obtain more texture compensation from different geometric decomposition domains.
\item Inspired by the second Nesterov acceleration with a fast convergence rate, we adopt a smartly chosen additional estimation rather than gradient evaluation to avoid complicated calculations as well as the risk of intermediate reconstruction results falling outside the geometric domains and ensure that the approximation results of intermediate stages are meaningful.
\item All the parameters in the proposed architecture are learnable end-to-end, and adaptive initializations are used to make the model flexible and ensure converging smoothly. Extensive experiments show that the proposed Nest-DGIL architecture outperforms other state-of-the-art methods.
\end{itemize}

\begin{figure*}[t]
\centerline{\includegraphics[width=2\columnwidth]
{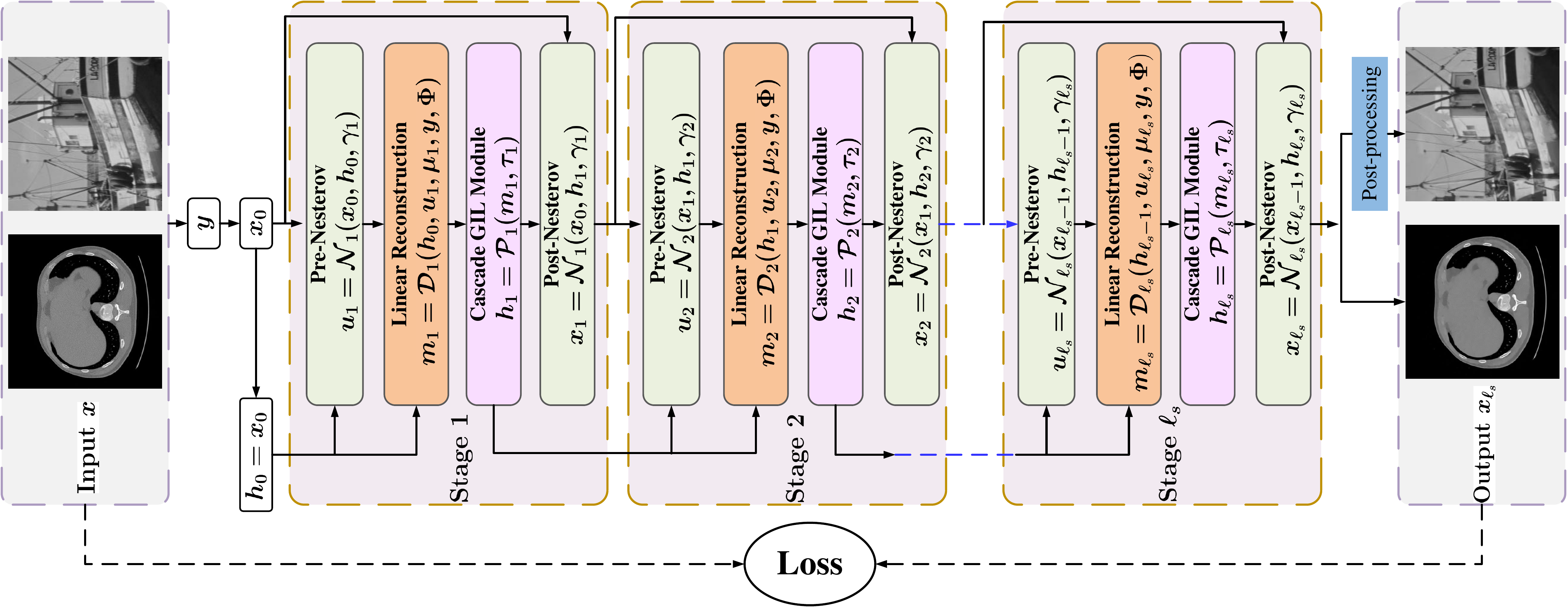}}
\caption{The overall architecture of our Nest-DGIL network. It consists of four main modules, i.e. linear reconstruction module $\mathcal{D}_{k}$, cascade geometric incremental learning module $\mathcal{P}_{k}$, Nesterov acceleration module $\mathcal{N}_{k}$ and post-processing module.}
\label{fig1}
\end{figure*}

The remainder of the paper is organized as follows. In Section \ref{secII},
after introducing the setup of image inverse problems and the algorithm of proximal gradient optimization, we present the second-Nesterov DGIL scheme to exactly impose
update accelerations by directly modifying the neural network
architecture. We then propose an operator spectral geometric decomposition approach to learn proximal-point solution in our framework. The experimental results are shown in Section \ref{secIII}. Finally,
we conclude the paper in Section \ref{secIV}.

\section{Methodology}\label{secII}
In this section, we propose an optimization-based image reconstruction, which is called an explainable Nest-DGIL framework (\emph{see} Fig. \ref{fig1}), and describe its specific formulation in detail. The proposed framework not only inherits the main advantages of the classical optimization approaches but also explicitly incorporates the process of embedding learnable nonlinear spectral geometric decomposition into a deep network.

\subsection{The second-Nesterov acceleration algorithm}
Before designing an explainable learning network for image reconstruction, we first analyze a proximal gradient-based reconstruction algorithm that satisfies the linear rate of convergence $(\mathcal{O}(\frac{1}{k}))$ \cite{Beck2009a}. We recall the image reconstruction problem \eqref{eq1a} as follow
\begin{equation}
\bm{x}=\underset{\bm{x}\in\mathcal{X}}{\arg \min }\left\{\mathcal{F}(\bm{x}):=\mathcal{S}(\bm{x})+\lambda\mathcal{R}(\bm{x})\right\}.
\label{eq3}
\end{equation}

{Based on the first-order gradient descent theory, 
we can obtain an approximation solution of \eqref{eq3} via the following alternative iterations
\begin{align}
\bm{m}_{k}&=\bm{x}_{k-1}-\mu_{k}\mathbf{\Phi}^{T}(\mathbf{\Phi} \bm{x}_{k-1}-\bm{y}),
\label{eq4-2}\\
\bm{x}_{k}&:=\underset{\bm{x}}{\arg \min } \left\{\frac{1}{2}\|\bm{x}-\bm{m}_{k}\|_{2}^{2}+\tau_k\mathcal{R}(\bm{x})\right\}, \label{eq5-2}
\end{align}
where $k$ denotes the iteration stage, $\mu_{k}=\frac{1}{\mathcal{L}_k}$ is the step-length, $\mathcal{L}_k$ is the estimated Lipschitz constant at stage $k$ and $\tau_{k}=\frac{\lambda}{\mathcal{L}_k}$ is the regularization parameter.} The proximal gradient-based reconstruction algorithm described above is generally considered to be a time consuming approach \cite{Boyd2010}. We can also see that it satisfies the linear rate of convergence $(\mathcal{O}(\frac{1}{k}))$ from the lemma \ref{lemma1} in the Appendix \ref{appendix-C}.

Let us now interpret the convergence rate of FISTA-type algorithms in a particular case that is representative in image inverse problems. FISTA which is a fast version of ISTA and adopts a well-designed linear combination of the previous reconstructions $\bm{x}_{k}$ and $\bm{x}_{k-1}$ as a refined acceleration step, has been proposed to solve \eqref{eq3} in an improved rate $(\mathcal{O}(\frac{1}{k^2}))$ \cite{Beck2009}. However, the result of a two-step update in FISTA has the risk of falling outside the geometric domains \cite{Liu2021}. Inspired by the second-Nesterov acceleration \cite{Nesterov1988,Liu2021}, we can approximate the solution of \eqref{eq3} via the following alternative iterations (denoted as Nesterov-II scheme)
\begin{align}
\bm{u}_{k}&
=\left(1-\gamma_{k}\right) \bm{x}_{k-1}+\gamma_{k} \bm{h}_{k-1},
\label{eq4-1}\\
\begin{split}
\bm{m}_{k}
&=\bm{h}_{k-1}-\mu_{k}\mathbf{\Phi}^{T}(\mathbf{\Phi} \bm{u}_{k}-\bm{y}),
\end{split}
\label{eq4}\\
\begin{split}
\bm{h}_{k}&
=\underset{\bm{x}}{\arg \min } \left\{\frac{1}{2}\|\bm{x}-\bm{m}_{k}\|_{2}^{2}+\tau_k\mathcal{R}(\bm{x})\right\}, \end{split}
\label{eq5}\\
\bm{x}_{k}&
=\left(1-\gamma_{k}\right) \bm{x}_{k-1}+\gamma_{k} \bm{h}_{k},
\label{eq5-1}
\end{align}
where $\gamma_{k}$ is a relaxation parameter.

The Nesterov II scheme \eqref{eq4-1}-\eqref{eq5-1} adopts a smartly chosen additional estimation rather than a gradient evaluation to avoid complicated calculations and achieves accelerated convergence $(\mathcal{O}(\frac{1}{k^2}))$. Furthermore, the Nesterov II scheme can ensure that the intermediate reconstruction results during iterations are all in the geometric domains \cite{Nesterov1988,Liu2021}.

\subsection{Operator spectral geometric decomposition}
One of the main ingredients in constructing an explainable neural network is to determine the learnable parameters by manually embedding the convolution and activation functions in an optimization-informed reconstruction model. The following technical decomposition is very useful for designing the learning module to solve the texture restoration problem \eqref{eq5}.
To extract features in more geometric domains, we consider an edge preservation regularizer defined by
$$
\mathcal{R}(\bm{x})=\sum_{\ell=1}^{s}\phi_\ell\left(\mathcal{K}(\boldsymbol{x})\right), \text { where } \phi_\ell(\bm{z})=g_\ell(\|\bm{z}\|)
$$
for some potential function $\phi_\ell:\mathbb{R}^d\mapsto\mathbb{R}$ defined in terms of some function $g_\ell: \mathbb{R} \mapsto \mathbb{R}$, and $\mathcal{K}$ is the feature extractor. However, the TV regularizer is a special case of our proposed edge-preserving regularizer $\mathcal{R}(\bm{x})$. We employ the edge-preserving regularizer $\mathcal{R}(\bm{x})$ to extract features in more geometric domains rather than only in the gradient domain. Taking the derivative of \eqref{eq5} with respect to $\bm{x}$, we can obtain
\begin{align}
\begin{split}
\bm{x}=&\bm{m}_{k}+\tau_k\frac{\partial\mathcal{R}(\bm{x})}{\partial\bm{x}}=\bm{m}_{k}+\tau_k\sum_{\ell=1}^{s}\mathcal{K}^{\prime}\phi^{\prime}_\ell\left(\mathcal{K}(\bm{x})\right)\\
=&\bm{m}_{k}+\mathcal{K}^{\prime}\psi\left(\mathcal{K}(\bm{x})\right)=\bm{m}_{k}+\mathcal{M}(\bm{x}),
\end{split}
\label{eq6}
\end{align}
where $\mathcal{M}(\bm{x})
$ denotes the nonlinear high-frequency geometric characteristic prior of image $\bm{x}$.

We can also understand the regularization operation (\ref{eq5}) as a compensating process to restore the missing texture $\bm{\omega}^*_{k}=\bm{x}^*-\bm{m}_k$ from the preliminary linear reconstruction $\bm{m}_{k}$. Since $\bm{x}$ is unknown, it is difficult to directly solve $\bm{x}$ from \eqref{eq6} with a nonlinear feature extractor $\mathcal{M}$. One of our goals is to devise numerical techniques that naturally account for this non-linearity by constructing an accurate and efficient scheme, which can admit physically feature correction series. The main technique we will use is operator spectral decomposition \cite{ANSELONE197467}, which is analyzed in Lemma 2 in Appendix \ref{appendix-C}.

The spectral radius of $\mathcal{M}(\bm{x})$ was mentioned to be dependent on the operator $\mathcal{K}$. If $\tau_k$ is small enough, the result in Lemma \ref{lemma2} easily shows that the nonlinear operator $\mathcal{M}$ satisfies the spectral constraint $\|\mathcal{M}\|<1$. Naturally, we can employ the Taylor expansion to split the nonlinear inverse operator $(I-\mathcal{M})^{-1}$ into different geometric domains for restoring $\bm{h}_{k}$, an approximation of the solution $\bm{x}$ in \eqref{eq6} is given as follows.
\begin{equation}
\begin{split}
\bm{h}_{k}&=(I-\mathcal{M})^{-1}(\bm{m}_{k})=\left(I+\sum_{i=1}^{\infty}\mathcal{M}^{i}\right)(\bm{m}_{k})\\&=\left(\sum_{i=0}^{n}\mathcal{M}^{i}+\mathcal{E}(\mathcal{M}^{n})\right)(\bm{m}_{k})=\bm{m}_{k}+\sum_{i=1}^{n+1}\bm{\omega}_{k,i},
\end{split}
\label{eq8}
\end{equation}
where $\bm{\omega}_{k,i}=\mathcal{M}^{i}(\bm{m}_{k})$ ($i=1,\dots,n$), $I=\mathcal{M}^{0}$, and $\mathcal{E}(\mathcal{M}^{n})(\cdot)=\theta_{\tau_k} \circ \mathcal{M}^{n}(\cdot)$ calculates the truncation remainder $\bm{\omega}_{k,n+1}=\mathcal{E}(\mathcal{M}^{n}(\bm{m}_{k}))$ of operator decomposition.

To proceed, let us define the left-hand and right-hand operators for any odd number $i$ ($i\leq n$) by (more explaination in Appendix \ref{appendix-A})
\begin{equation}
\begin{split}
\bm{m}_{k,0}
=\bm{m}_{k},\;&\cdots,\;
\bm{m}_{k,i}=\psi\left(\mathcal{K}(\bm{m}_{k,i-1})\right),\\
\bm{h}_{k}=\bm{r}_{k,0}+\bm{m}_{k,0},\;&\cdots,\;\bm{r}_{k,i-1}=\mathcal{K}^{\prime}(\bm{h}_{k,i}),
\end{split}
\label{eq7}
\end{equation}
otherwise for any even number $i$ ($i\leq n$), one has
\begin{equation}
\begin{split}
\bm{m}_{k,0}
=\bm{m}_{k},\;&\cdots
,\;\bm{m}_{k,i}=\mathcal{K}^{\prime}(\bm{m}_{k,i-1}),\\
\bm{h}_{k}=\bm{r}_{k,0}+\bm{m}_{k,0},\;&\cdots,\;
\bm{r}_{k,i-1}=\psi\left(\mathcal{K}(\bm{h}_{k,i})\right),
\end{split}
\label{eq7a}
\end{equation}
where $\bm{h}_{k,i}=\bm{m}_{k,i}+\bm{r}_{k,i}$.
Thus we rewrite $\bm{\omega}_{k,i}\in \mathcal{X}_i$ ($i\leq n$) and $\bm{\omega}_{k,n+1}$ as follows
\[\bm{\omega}_{k,i}=\mathcal{M}^{i}(\bm{m}_{k})=\underbrace{\underline{\mathcal{K}^{\prime}\psi(\mathcal{K}}[\cdots\underline{\mathcal{K}^{\prime}\psi(\mathcal{K}}[\underline{\mathcal{K}^{\prime}\psi(\mathcal{K}}}_{i}(\bm{m}_{k}))])],\]
\[\bm{\omega}_{k,n+1}=\left\{\begin{array}{cl}
\underbrace{\underline{\mathcal{K}^{\prime}}\underline{\psi(\mathcal{K}}[\cdots\underline{\psi(\mathcal{K}[}}_{n}\underline{\mathcal{K}^{\prime}\bm{\theta}_{\tau_k}\big(\psi(\mathcal{K}}(\bm{m}_{k,n}))])\big)]),\qquad & \\
\hfill\text{ if } n \text{ is even number}; &\\
\underbrace{\underline{\mathcal{K}^{\prime}}\underline{\psi(\mathcal{K}}[\cdots\underline{[\mathcal{K}^{\prime}}}_{n}\underline{\psi(\mathcal{K}\bm{\theta}_{\tau_k}\big(\mathcal{K}^{\prime}}(\bm{m}_{k,n}))]\big)]),\qquad  & \\
\hfill\text{ if } n \text{ is odd number}. &
\end{array}\right.\]
where $\mathcal{X}_i$ is a geometric characteristic subspace, and image space $\mathcal{X}$ is a linear combination of subspace $\{\mathcal{X}_i\}_{i=1}^{n+1}$.
Finally we can define the geometric incremental component $\bm{\omega}_{k}$ by
\begin{equation}
\bm{\omega}_{k}=\bm{h}_{k}-\bm{m}_{k}=\sum_{i=1}^{n}\mathcal{M}^{i}(\bm{m}_{k})+\mathcal{E}(\mathcal{M}^{n}(\bm{m}_{k}))=\sum_{i=1}^{n+1}\bm{\omega}_{k,i},
\end{equation}
where $\mathcal{M}^{i}(\bm{m}_{k})$ represents the texture compensation from the decomposition of the features $\bm{\omega}_{k,i}$ in subspace $\mathcal{X}_i$. We also list the spectral geometric decomposition schemes of two examples $\phi_\ell(r)=\frac{1}{2}\|r\|^{2}$ and $\phi_\ell(r)=\|r\|$ in Appendix \ref{appendix-B}.

\begin{figure*}[htbp]
\centerline{\includegraphics[width=1.8\columnwidth]
{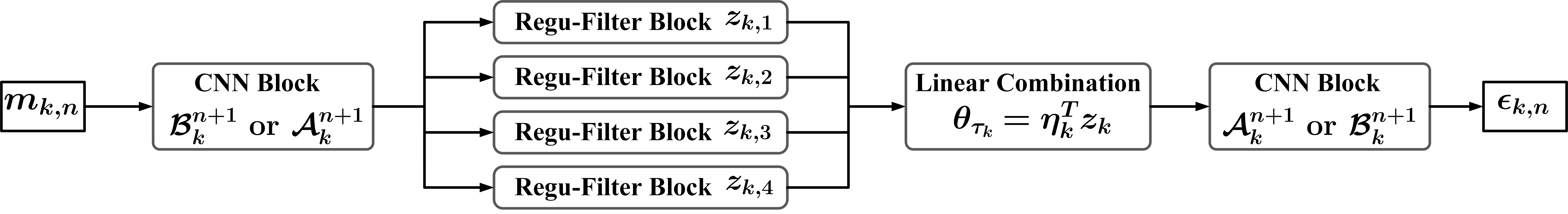}}
\caption{Spectral decomposition truncated error ${\epsilon_{k,n}}.$}
\label{fig2_Geometric_Increment_Block_b}
\end{figure*}

\subsection{Truncation optimization problem}
In this part, we aim to present an optimization minimization to model high-frequency feature restoration, which can help to obtain the remainder of the truncation $\mathcal{E}(\mathcal{M}^{n}(\bm{m}_{k}))$.

We have extracted the decomposed principal high-frequency feature component $\sum_{i=1}^{n}\bm{\omega}_{k,i}$, then we will approximate the truncation information $\bm{\omega}_{k,n+1}$ by solving the optimization with $\mathcal{R}(\bm{x})=\|\bm{x}\|_{p}$ in \eqref{eq5}, and the solution is rewritten as
\begin{equation}
\bm{x}=\text{Prox}_{\tau_{k}\|\cdot\|_{p}}(\bm{z}^0_k)=\underset{\bm{x}}{\arg \min } \frac{1}{2}\|\bm{x}-\bm{z}^0_k\|_{2}^{2}+\tau_{k}\|\bm{x}\|_{p},
\label{eq10}
\end{equation}
where $\bm{z}^0_k=\psi(\mathcal{K}(\bm{m}_{k,n}))$ for an even number $n$ and $\bm{z}^0_k=\mathcal{K}^\prime(\bm{m}_{k,n})$ for an odd number $n$.
Relying only on the $\ell_{1}$ regularization prior is not enough to reconstruct geometric features from different abundant texture and smooth regions, thus we propose a better truncation remainder by different regularisers $\|\bm{z}\|_{p}$ with different values of $p$ to extract more useful information from the remainder term. In fact, problem \eqref{eq10} has a closed form solution for the given $p$-values \cite{Bauschke2011,Chaux2007,Combettes2008}, that is,
\begin{equation*}
\begin{split}
\bm{z}_{k,1}&=\text{Prox }_{\tau_{k}\|\cdot\|_0}\left(\bm{z}^0_k\right)=\left\{\begin{array}{cc}
\bm{z}^0_k, & |\bm{z}^0_k|\geq{(2\tau_{k})}^{1/2} \\
0, & \text{ otherwise }
\end{array}\right.;\\
\bm{z}_{k,2}&=\text{Prox }_{\tau_{k}\|\cdot\|_1}\left(\bm{z}^0_k\right)=\operatorname{sign}(\bm{z}^0_k) \max\{|\bm{z}^0_k|-\tau_{k},0\};\\
\bm{z}_{k,3}&=\text{Prox }_{\tau_{k}\|\cdot\|_{\frac{3}{2}}}\left(\bm{z}^0_k\right)\\&=\bm{z}^0_k+\frac{9}{8} \tau_{k}^{2} \operatorname{sign}(\bm{z}^0_k)\left(1-\sqrt{1+\frac{16|\bm{z}^0_k|}{9 \tau_{k}^{2}}}\right);\\
\bm{z}_{k,4}&=\text{Prox }_{\tau_{k}\|\cdot\|_2}\left(\bm{z}^0_k\right)=\frac{\bm{z}^0_k}{1+2\tau_{k}}.
\end{split}
\label{eq11}
\end{equation*}

By a weighted sum of different shrinkage-thresholding results, we can obtain the update scheme of the geometric incremental compensation \eqref{eq5} as
\begin{equation*}
\begin{split}
\bm{h}_{k}=&\bm{m}_{k}+\sum_{i=1}^{n}\underbrace{\underline{\mathcal{K}^{\prime}\psi(\mathcal{K}}[\cdots\underline{\mathcal{K}^{\prime}\psi(\mathcal{K}}[\underline{\mathcal{K}^{\prime}\psi(\mathcal{K}}}_{i}(\bm{m}_{k}))])]\\
&+\left\{\begin{array}{c}
\underbrace{\underline{\mathcal{K}^{\prime}}\underline{\psi(\mathcal{K}}[\cdots\underline{\psi(\mathcal{K}[}}_{n}\underline{\mathcal{K}^{\prime}\bm{\theta}_{\tau_k}\big(\psi(\mathcal{K}}(\bm{m}_{k,n}))])\big)]),\\
\hfill\text{ if } n \text{ is even number}; \\
\underbrace{\underline{\mathcal{K}^{\prime}}\underline{\psi(\mathcal{K}}[\cdots\underline{[\mathcal{K}^{\prime}}}_{n}\underline{\psi(\mathcal{K}\bm{\theta}_{\tau_k}\big(\mathcal{K}^{\prime}}(\bm{m}_{k,n}))]\big)]), \\
\hfill\text{ if } n \text{ is odd number},
\end{array}\right.
\end{split}
\label{eq_gil_15}
\end{equation*}
where $\theta_{\tau_k}(\bm{z}^0_k)=\sum_{i=1}^4\eta_{k,i} \bm{z}_{k,i}
$, $\eta_{k,i}=\beta_{k,i}/\sum_{i=1}^4\beta_{k,i}$ is the normalization for learnable parameters $\beta_{k,i}$ ($i=1,\dots,4$) that adapts different importance of every regulariser.

\subsection{Nesterov-informed DGIL framework}
As suggested in the previous section, a new designed framework will not only have geometric decomposition properties similar to the above classical Nesterov-based optimization, but the new approach will also be better in explaining the analysis for the proposed network. In this part, we should analyze four main components of the proposed Nest-DGIL framework in detail, including a linear reconstruction module $\mathcal{D}_{k}$, a cascade GIL module $\mathcal{P}_{k}$, Nesterov acceleration module $\mathcal{N}_{k}$, and loss function.

\textbf{Linear reconstruction module $\mathcal{D}_{k}$:}
The module $\mathcal{D}_{k}$ corresponding to \eqref{eq4} is used directly to generate the preliminary approximation solution $\bm{m}_{k}$, which can be denoted by
\[\mathcal{D}_{k}(\bm{h}_{k-1},\bm{u}_k,\mu_k,\bm{y},\boldsymbol{\Phi})=\bm{h}_{k-1}-\mu_{k}\mathbf{\Phi}^{T}(\mathbf{\Phi} \bm{u}_{k}-\bm{y}).\]
It is well known that the step length $\mu_{k}$ should be positive and decreases with the increase of iterations smoothly. To increase the flexibility of the network, we set the step length $\mu_{k}$ to be learnable during iterations, while it is fixed in traditional model-based methods. There are a variety of ways to use training data to adaptively learn step length $\mu_{k}$. To facilitate backpropagation, we employ the \textbf{softplus} function $\text{\emph{sp}}(x)=\ln (1+\exp (x))$ to implement the initialization of the learnable step length $\mu_{k}$ \cite{Xiang2020}, and the initial guess for the stage $k$ is given by
\begin{equation}
\mu_{k}=\text{\emph{sp}}\left(\alpha_{1} k+c_{1}\right), \alpha_{1}<0, k=1,2, \ldots, \ell_{s}.
\label{eq_mu}
\end{equation}

\textbf{Cascade GIL module $\mathcal{P}_{k}$:}
We also notice that module $\mathcal{D}_{k}$ often results in heavy artifacts. Furthermore, we design a cascade GIL module
$\bm{h}_k=\mathcal{P}_{k}(\bm{m}_k,\tau_k)$
 to extract the high frequency feature defined in different geometric characteristic domains $\mathcal{X}_i$.

The geometric texture information in each domain $\mathcal{X}_i$
can be represented by the partial derivatives of image $\bm{x}$, \emph{e.g.}, feature extractor $\mathcal{K}=\nabla$ or $\mathcal{K}=\Delta$. It is well known that the convolution in neural network can be seen as the combination of several derivative operations.
So we start by treating the operator $\mathcal{K}$ and function $\psi(\cdot)$ as an embedded convolution $Conv(\cdot)$ and a composited Rectified Linear Unit (ReLU) activation. Therefore, the operation $\mathcal{A}^i_{k}(\cdot)$ in each cascade of the GIL module $\mathcal{P}_{k}$ is given by the new network layer $\text{ReLU}\left(Conv(\cdot)\right)$ to replace a non-linear operator $\psi\left(\mathcal{K}(\bm{x})\right)$. Inversely, since the intensities of the approximation image $\bm{h}_{k,i}$ processed by $\mathcal{K}^{\prime}(\cdot)$ are non-negative, we filter out the negative values in $Conv(\bm{h}_{k,i})$ with ReLU without changing the non-negative values. Similar to $\mathcal{A}^i_{k}(\cdot)$, we design the layer $\mathcal{B}^i_{k}(\cdot)=\text{ReLU}\left(Conv(\cdot)\right)$ to represent $\mathcal{K}^{\prime}(\cdot)$. We believe that this is reasonable, 
which means that clustering these cascades directly corresponds to finding all geometric feature compensations that are dominated by the same stage. In summary, the proposed GIL module $\mathcal{P}_{k}$ strictly corresponds to the geometric spectral decomposition of the operator.

\begin{figure}[htbp]
\centerline{\includegraphics[width=1\columnwidth]
    {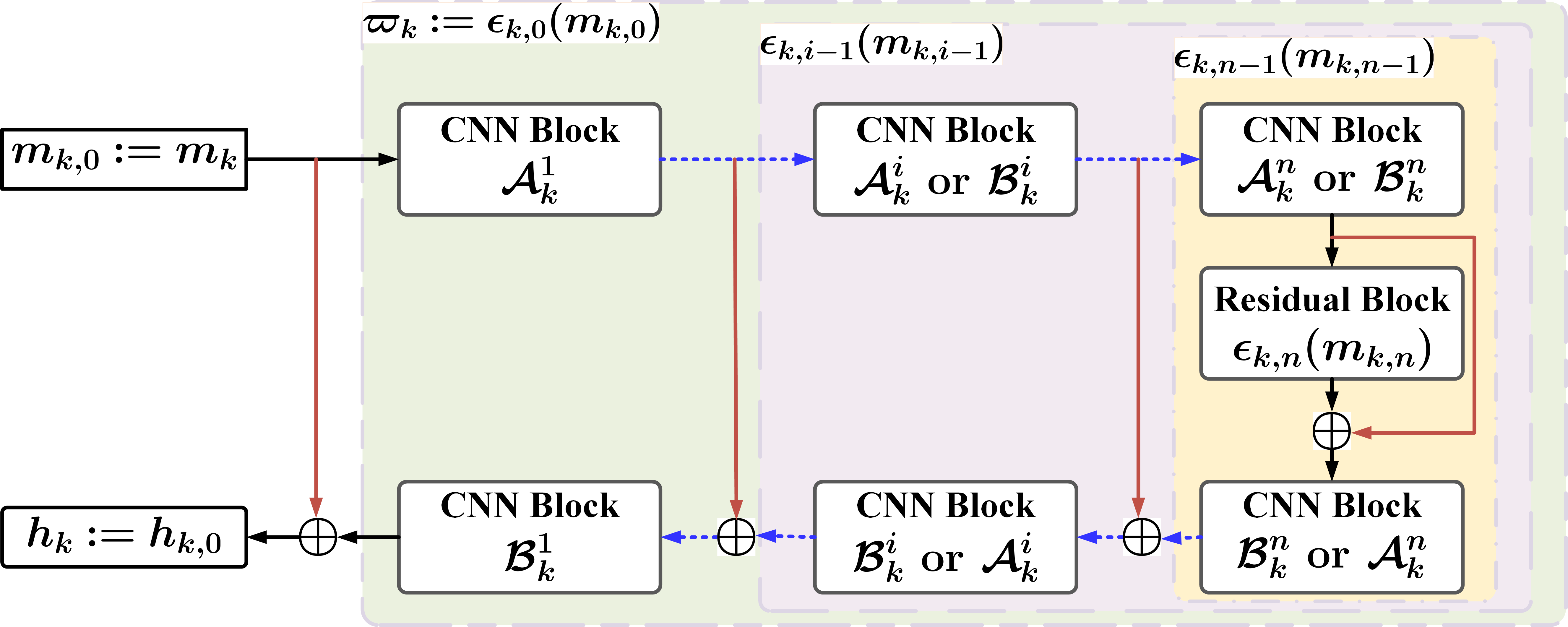}}
\caption{{Cascade GIL module ${h_{k}=\mathcal{P}_{k}(m_k,\theta_k)=m_k+\varpi_k}$ and geometric incremental component ${\varpi_{k}}$, where the module shares same parameter $\mathcal{A}_k^i$ and $\mathcal{B}_k^i$ in each ${\varpi_{k,i}}$.}}
\label{fig2_Geometric_Increment_Block_d}
\end{figure}

\begin{figure}[htbp]
\centerline{\includegraphics[width=1\columnwidth]
{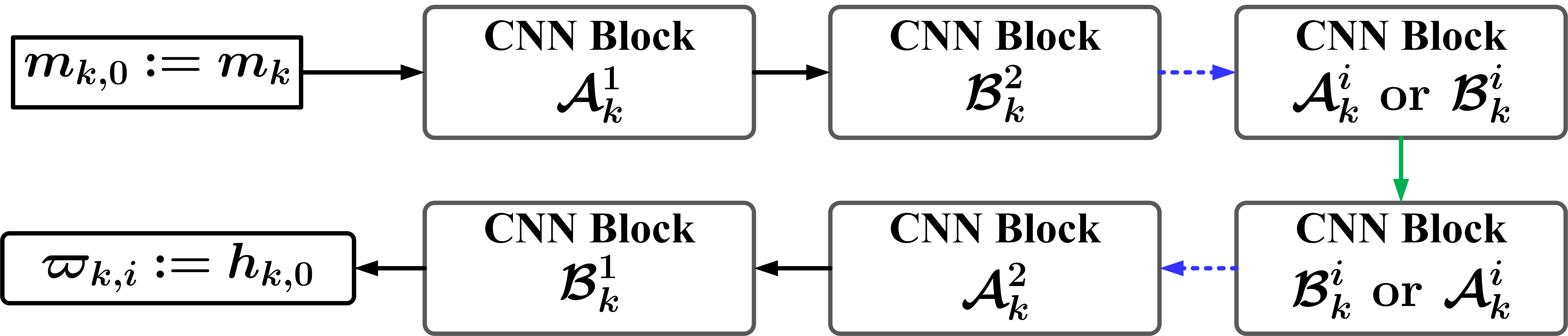}}
\caption{Spectral decomposition component  ${\varpi_{k,i}}$ in the geometric subspace ${\mathcal{X}_i\;(1\leq i\leq n)}.$}
\label{fig2_Geometric_Increment_Block_a}
\end{figure}

To proceed, we first construct the final incremental error by (\emph{see} Fig. \ref{fig2_Geometric_Increment_Block_b})
\begin{equation*}
\bm{\epsilon}_{k,n}(\bm{m}_{k,n})=\left\{\begin{array}{ll}
\mathcal{B}^{n+1}_{k}(\theta_{\tau_k}(\mathcal{A}^{n+1}_{k}(\bm{m}_{k,n}))),\\
\hfill\text{ if } n \text{ is even number};\\
\mathcal{A}^{n+1}_{k}(\theta_{\tau_k}(\mathcal{B}^{n+1}_{k}(\bm{m}_{k,n}))),\\
\hfill\text{ if } n \text{ is odd number}.
\end{array}\right.
\label{eq14-1}
\end{equation*}
Consequently, we are now going to show an incremental learning extrapolation relationship from \eqref{eq7} to expand the cascade architecture by
\begin{equation*}
\begin{split}
\bm{m}_{k,i}&+\bm{\epsilon}_{k,i}(\bm{m}_{k,i})\\&=\left\{\begin{array}{cl}
\bm{m}_{k,i}+\mathcal{B}^{i+1}_{k}\left(\bm{m}_{k,i+1}+\bm{\epsilon}_{k,i+1}(\bm{m}_{k,i+1})\right),\; & \\
\hfill\text{ if } i \text{ is even number}; &\\
\bm{m}_{k,i}+\mathcal{A}^{i+1}_{k}\left(\bm{m}_{k,i+1}+\bm{\epsilon}_{k,i+1}(\bm{m}_{k,i+1})\right),\;& \\
\hfill\text{ if } i \text{ is odd number}, &
\end{array}\right.
\end{split}
\label{eq14}
\end{equation*}
where $i=n-1,\cdots,1,0$.

\begin{figure}[h]
\centerline{\includegraphics[width=1\columnwidth]
{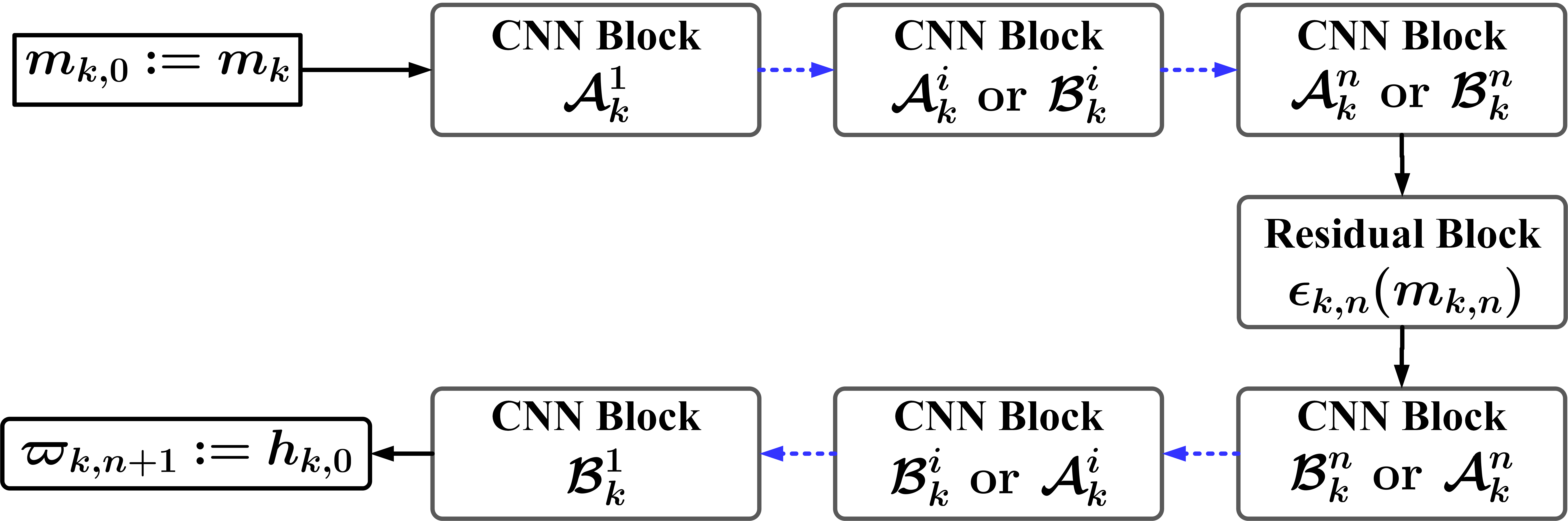}}
\caption{Spectral decomposition truncated component ${\varpi_{k,n+1}}$ in the geometric subspace ${\mathcal{X}_{n+1}}.$}
\label{fig2_Geometric_Increment_Block_c}
\end{figure}

The total incremental learning estimation in the stage $k$ is performed in Fig.\ref{fig2_Geometric_Increment_Block_d} by
\begin{equation}
\begin{split}
\bm{h}_{k}
=&\mathcal{P}_{k}(\bm{m}_k,\tau_k)
=\bm{m}_{k}+\varpi_k\\
=&\bm{m}_{k}+(\sum_{i=1}^{n}\varpi_{k,i}+ \varpi_{k,n+1})
\\
=&\bm{m}_{k}+\sum_{i=1}^{n}\underbrace{\underline{\mathcal{B}^{1}_k\mathcal{A}_k^2}[\underline{\mathcal{B}^3_k\cdots}\underline{\cdots\mathcal{A}^3_k}[\underline{\mathcal{B}^{2}_k\mathcal{A}^1_k}}_{i}(\bm{m}_{k})]]\\
+&
\left\{\begin{array}{c}
\underbrace{\underline{\mathcal{B}^1_k}\;\underline{\mathcal{A}^2_k}[\cdots\underline{\mathcal{A}^n_k[}}_{n}\underline{\mathcal{B}^{n+1}_k\bm{\theta}_{\tau_k}\big(\mathcal{A}^{n+1}_k}(\bm{m}_{k,n})])\big)],\\
\hfill\text{ if } n \text{ is even number}; \\
\underbrace{\underline{\mathcal{B}^1_k}\;\underline{\mathcal{A}^2_k}[\cdots\underline{\mathcal{B}^n_k[}}_{n}\underline{\mathcal{A}^{n+1}_k\bm{\theta}_{\tau_k}\big(\mathcal{B}^{n+1}_k}(\bm{m}_{k,n})])\big)], \\
\hfill\text{ if } n \text{ is odd number}.
\end{array}\right. 
\end{split}
\label{eq21}
\end{equation}
{where $\varpi_{k,i}$-block and $\varpi_{k,n+1}$-block strictly corresponding to the operator spectral geometric decomposition \eqref{eq_gil_15} are denoted in Fig. \ref{fig2_Geometric_Increment_Block_a} and Fig. \ref{fig2_Geometric_Increment_Block_c}.}

Now the task of finding an optimal Nest-DGIL framework
such that its prediction $\bm{x}_{\ell_s}$ is similar to ground truth $\bm{x}$ is to learn the network parameters and physical hyperparameters.
To proceed, we give the concrete convolution sets of $\mathcal{A}={\{\mathcal{A}_k^i\}}^{\ell_s,n+1}_{k,i=1}$ and $\mathcal{B}={\{\mathcal{B}_k^i\}}^{\ell_s,n+1}_{k,i=1}$ in the proposed cascade GIL module. Each $\mathcal{A}_k^i$ ($2\leq i\leq n+1$) corresponds to $N_{f}$ filters where each filter is of size $3\times3\times N_{f}$, while $\mathcal{A}_k^{1}$ corresponds to $N_{f}$ filters where each filter is of size $3\times3$. The operation $\mathcal{B}_{k}^{1}$ corresponds to 1 filter of size $3 \times 3 \times N_{f}$, but the other $\mathcal{B}_{k}^{i}$ corresponds to $N_{f}$ filters (each filter is of size $3 \times 3 \times N_{f}$). All the CNN blocks ($\mathcal{A}_{k}^{i}$ and $\mathcal{B}_{k}^{i}$) contain ReLU activation except the nearest one before and after the residual block. The last CNN block without ReLU before the residual block can make full use of the previous input information for truncation optimization with multi-regularisers. The first CNN block without ReLU after the residual block can comprehensively utilize the preliminary linear reconstruction and the geometric incremental information learned from the proposed cascade incremental learning module.

Based on prior knowledge, the threshold value $\tau_k$ should be positive and decrease with increasing iterations smoothly. Similar to module $\mathcal{D}_{k}$, the initial guess of learnable $\tau_k$ is given as follows
\begin{equation}
\tau_k=\text{\emph{sp}}\left(\alpha_{2} k+c_{2}\right),\alpha_{2}<0,k=1,2, \ldots, \ell_{s}.
\label{eq33}
\end{equation}

Unfortunately, we notice that batch normalization (BN) is not adopted in many reconstruction networks because some recent papers showed that the BN layer is more likely to yield undesirable results when a network becomes deeper and more complex \cite{Wang2018,Zhang2018a}.

\begin{algorithm}[t]
    \caption{{The Proposed Nest-DGIL Framework.}}
    \label{algorithm1}

    {\bf Input:} The measurement matrix $\mathbf{\Phi}$ and its transpose $\mathbf{\Phi}^{T}$, the geometric incremental domain number $n+1$, the stage number $\ell_s$, the training dataset $\mathcal{T}=\left\{\left(\bm{y}^{i}, \bm{x}^{i}\right)|i=1,\cdots,N_b\right\}$.\\
    {\bf Initialize:} $\bm{x}_0=\mathbf{\Phi}^{T}\bm{y}$, $\bm{h}_0=\bm{x}_0$, the learnable parameters $\Theta=\left\{\mu_{k}, \tau_k, \gamma_{k}, \mathcal{A}^i_k(\cdot), \mathcal{B}^i_k(\cdot),\beta_{k,j}\right\}_{k,i,j=1}^{\ell_{\mathrm{s}},n+1,4}$. \\
    {\bf Inference:} 
    \begin{algorithmic}[1]
        \For{$k=1,2,\cdots,\ell_s$}
        \State $\bm{u}_{k}=\left(1-\gamma_{k}\right) \bm{x}_{k-1}+\gamma_{k} \bm{h}_{k-1}$; // \eqref{eq4-1}
        \State  $\bm{m}_{k}=\bm{h}_{k-1}-\mu_{k}\mathbf{\Phi}^{T}(\mathbf{\Phi} \bm{u}_{k}-\bm{y})$; // \eqref{eq4}
        \State  $\bm{h}_{k}=\bm{m}_{k}+(\sum_{i=1}^{n}\varpi_{k,i}+ \varpi_{k,n+1})$; // \eqref{eq21}
        \State $\bm{x}_{k}=\left(1-\gamma_{k}\right) \bm{x}_{k-1}+\gamma_{k} \bm{h}_{k}$ // \eqref{eq5-1}
        \EndFor
        \While{CS reconstruction based on patch-by-patch}
        \State Weighted reconstruction of the overlapping patches;
        \EndWhile
    \end{algorithmic}
    {\bf Training:}
    \begin{algorithmic}[1]
        \State $\mathcal{L}_{\mathrm{total}}=\frac{1}{N_{b} N} \sum_{i=1}^{N_{b}}\left\|\bm{x}^{i}_{\ell_s}-\bm{x}^{i}\right\|_{1};$ // \eqref{eq23}
    \end{algorithmic}
    {\bf Output:}
    $\mathcal{H}(\mathcal{T} ; \Theta)=\bm{x}_{\ell_s}.$
\end{algorithm}

\textbf{Nesterov acceleration module $\mathcal{N}_{k}$:} The Nesterov acceleration module
\[\mathcal{N}_{k}(\bm{x}_{k-1},\bm{h}_{k-1},\gamma_{k})=\left(1-\gamma_{k}\right) \bm{x}_{k-1}+\gamma_{k} \bm{h}_{k-1}\]
 uses previous reconstruction result to correct the current descending direction and realizes that the acceleration of reconstruction convergence corresponds to \eqref{eq4-1} and \eqref{eq5-1}. Based on the prior knowledge, the relaxation parameters $\gamma_{k}$ and $1-\gamma_{k}$ should be positive, and $\gamma_{k}$ increases with the increase in iterations smoothly. We design the initial guess of learnable step-length $\gamma_{k}$ as follows
\begin{equation}
\begin{split}
\gamma_{k}^{\prime}&=\text{sp}\left(\alpha_{3} k+c_{3}\right),\alpha_{3}>0,\\
\gamma_{k} &=\sigma(\gamma_{k}^{\prime})=\frac{e^{\gamma_{k}^{\prime}}}{e^{1-\gamma_{k}^{\prime}}+e^{\gamma_{k}^{\prime}}}, k=1,2, \ldots, \ell_{s},
\end{split}
\label{eq_gamma_all}
\end{equation}
where the sigmoid function $\sigma(\cdot)$ ensures that $\gamma_{k}$ and $1-\gamma_{k}$ are positive.

Finally, the overall Nest-DGIL algorithm is summarized in Algorithm \ref{algorithm1}, and the overall network framework is shown in Fig.\ref{fig1}. Throughout all experiments, we notice that  Nest-DGIL and Nest-DGIL+ denote the proposed framework with a fixed random Gaussian sampling matrix and a jointly learned sampling matrix, respectively. In addition, the same weighted reconstruction postprocessing technique of overlapping patches as CSformer \cite{Ye2023} is also performed in the Nest-DGIL and Nest-DGIL+ frameworks.

\begin{figure*}[htbp]
    \centering
    \includegraphics[width=1\textwidth]{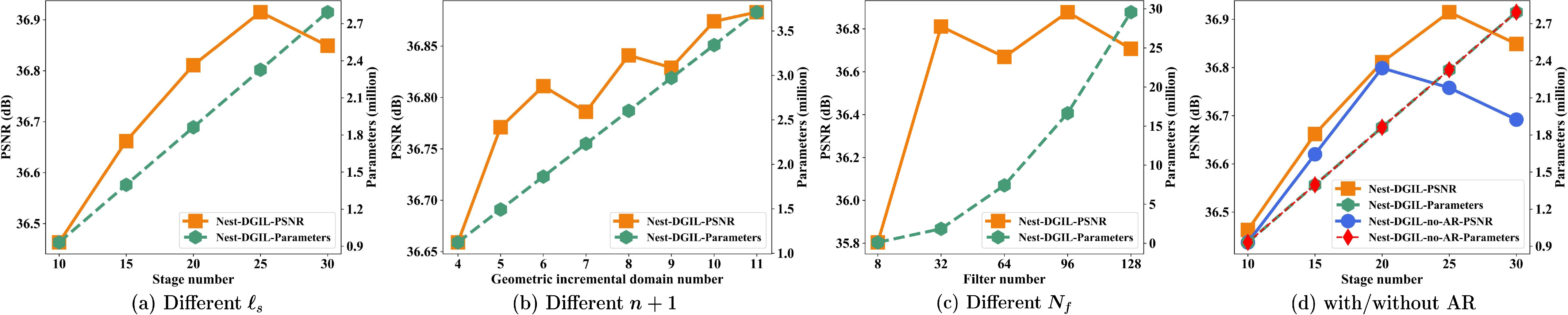}
    \caption{Evaluations of the network in Set11 with the CS ratio 25\%.}
    \label{fig_net_parameters}
\end{figure*}

\textbf{Loss function:}
We can achieve reconstruction prediction $\bm{x}_{\ell_s}^{i}$ from the proposed Nest-DGIL framework trained using the training data pairs $\left\{\left(\bm{y}^{i}, \bm{x}^{i}\right)|i=1,\cdots,N_b\right\}$. The loss function is commonly used to seek the optimal network parameters. Here we minimize the loss function between the final reconstruction output $\bm{x}_{\ell_s}$ and the ground truth $\bm{x}$, which is formulated as follows
\begin{equation}
\mathcal{L}_{\mathrm{total}}=\frac{1}{N_{b} N} \sum_{i=1}^{N_{b}}\left\|\bm{x}^{i}_{\ell_s}-\bm{x}^{i}\right\|_{1}.
\label{eq23}
\end{equation}
where $N_{b}$ and $N$ denote the number of data-pairs and the size of each image, respectively. A pixel-based $\ell_{1}$-loss is used to extract perceptually high-frequency geometric details \cite{Jiang2018}.

\textbf{Parameters and initialization:}
Four modules in every stage of the proposed framework strictly correspond to the Nesterov II scheme from \eqref{eq4-1} to \eqref{eq5-1}. Learnable parameters $\Theta=\left\{\mu_{k}, \tau_k, \gamma_{k}, \mathcal{A}^i_k(\cdot), \mathcal{B}^i_k(\cdot),\beta_{k,j}\right\}_{k,i,j=1}^{\ell_{\mathrm{s}},n+1,4}$ consist of step length $\mu_{k}$, relaxation parameter $\gamma_{k}$, regularization parameter $\tau_k$, convolutional layers $\mathcal{A}^i_k(\cdot)$ and $\mathcal{B}^i_k(\cdot)$, weight coefficient $\beta_{k,j}$. All these parameters are learned as neural network parameters by minimizing the loss \eqref{eq23}.

Similarly to the conventional model-based method, the proposed method also requires an initial input $\bm{x}_0$ from a given $\bm{y}$ in Fig. \ref{fig1}. For natural image CS and sparse-view CT, we employ $\bm{x}_0=\mathbf{\Phi}^{T}\bm{y}$ for initialization. We use the initialization $\bm{h}_0=\bm{x}_0$ for both natural-image CS and sparse-view CT. The convolution network is initialized with the Kaiming initialization \cite{He2015}. $\{\alpha_{1}, \alpha_{2}, \alpha_{3}\}$ and $\{c_{1}, c_{2}, c_{3}\}$ are initialized with $\{-0.2, -0.5,0.5\}$ and $\{0.1, -2, 0\}$, respectively. Throughout all experiments, the weight coefficients $\{\beta_{k,j}\}_{k,j=1}^{\ell_{\mathrm{s}},4}$ are initialized with $\{1, 1, 1, 1\}$ at all stages. The geometric incremental domain number $n+1$ and filter number $N_{f}$ are set as $6$ and $32$, the stage number $\ell_s$ is set as $20$ for natural image CS and $7$ for sparse view CT, respectively. Following the CSformer configurations \cite{Ye2023}, the overlap step is also set to 8 in our weighted reconstruction of the overlapping patches.

\section{Experimental results and discussion}\label{secIII}
We compare the proposed method with the state-of-the-art methods on two representative CS tasks (natural image CS and sparse-view CT). Three measures including Peak Signal-to-Noise Ratio (PSNR), Structural Similarity Index Measure (SSIM), and Root Mean Square Error of Hounsfield Unit (RMSE(HU)) are employed to evaluate their reconstruction performances.

\subsection{Implementation details}
For natural image CS reconstruction, we utilize the 400 training images of size 180 $\times$ 180 \cite{Chen2017a} to generate the training data pairs $\left\{\left(\bm{y}^{i}, \bm{x}^{i}\right)\right\}_{i=1}^{N_{b}}$ with size 33 $\times$ 33 \cite{You2021}. Meanwhile, we increase the diversity of training data by applying the data augmentation technique \cite{You2021}. We employ widely used benchmark datasets Set11\footnote{http://dsp.rice.edu/software/DAMP-toolbox.}, BSD68 \cite{Martin2001} and Urban100 \cite{Huang2015} for image CS reconstruction testing. The CS ratios $\{10\%, 25\%, 30\%, 40\%, 50\%\}$ are evaluated in our natural image experiments. We apply $\bm{y}^{i}=\mathbf{\Phi} \bm{x}^{i}$ to yield CS measurements like \cite{You2021}, where $\bm{x}^{i}$ is the vectorized image block and $\mathbf{\Phi} \in \mathbb{R}^{M \times N}$ is a random Gaussian measurement matrix orthogonalizing its rows ($\mathbf{\Phi}\mathbf{\Phi}^{T}=I$, $I$ is the identity matrix) in every CS ratio. When $\mathbf{\Phi}$ is jointly learned, we utilize a convolution layer to mimic the CS sampling process $\bm{y}^{i}=\mathbf{\Phi} \bm{x}^{i}$ and another convolution layer with $N$ filters from $\mathbf{\Phi}^{T} \in \mathbb{R}^{N \times M}$ to obtain initialization $\mathbf{\Phi}^{T}\bm{y}$ as \cite{Zhang2020}.

Due to not only the frequency of requests by Mayo and AAPM for the complete 2016 Grand Challenge dataset but also the lack of such data,
therefore, investigators at the Mayo Clinic build a library of CT patient projection data named "Low Dose CT Image and Projection Data" \cite{Moen2020}.
This unique data library has facilitated the development and validation of new CT reconstruction algorithms. For sparse-view CT, we chose the first ten-patient dataset of noncontrast chest CT scans to evaluate the reconstruction performances of the compared methods, which contains 3324 full-dose CT images of 1.5 mm thickness. Among them, the first eight-patient subset is used for training and the ninth patient's data for validation whereas the last one for testing. In detail, there are 2621 slices of 512 $\times$ 512 images for training and 340 slices of 512 $\times$ 512 images for validation. 363 slices of 512 $\times$ 512 size are used as LDCT-Data testing dataset. For a sparse view CT generalizability test, we employ patient data from the "Fused Radiology-Pathology Lung Dataset" \cite{Rusu2017}, which has 321 slices of size 512 $\times$ 512 with 1 mm thickness as the FRPLung-Data testing dataset. The training images are augmented by performing horizontal and vertical flips. Sinograms for this dataset are 729 pixels by 720 views, and the original artifact-free images are reconstructed using the iradon transform in TorchRadon \cite{Ronchetti2020} using all 720 views. The projection observations are down-sampled to 60, 90, 120 and 180 views to simulate a few view geometries, respectively.

We use Pytorch to implement the proposed Nest-DGIL framework with a batch size of 32 for natural image CS and a batch size of 1 for sparse view CT separately. We use Adam optimization \cite{Kingma2014} with a learning rate of 0.0001 to train the network. In natural image reconstruction, 200 epochs are used, whereas in CT reconstruction 50 epochs are used. All experiments are performed on a workstation with Intel Xeon CPU E5-2630 and Nvidia Tesla V100 GPU.

\subsection{Intra-method evaluation}
In this part, through several groups of evaluations, we aim to investigate the role of different network components in our Nest-DGIL in the reconstruction performance of CS in natural images, including stage number $\ell_s$, geometric incremental domain number $n+1$, filter number $N_{f}$ and adaptive remainder optimization (AR) \eqref{eq_gil_15}, as well as ablation study of the proposed cascade incremental learning module (CI) \eqref{eq21}, AR and deep Nesterov II scheme (N-II) (\eqref{eq4-1} to \eqref{eq5-1}), and different shared settings of Nest-DGIL.

\textbf{Test of stage number $\ell_s$:}
To better explore the reconstruction performance of the proposed method, we analyze the reconstruction results by varying the stage number $\ell_s$ from 10 to 30 at 5 intervals. Using CNN architectures with different stage numbers $\ell_s$ in Set11 with CS ratio 25\%, the average PSNR curves of the reconstructed results are shown in Fig.\ref{fig_net_parameters}(a). We can observe that the average PSNR curves increase as $\ell_s$ increases and become stable after $\ell_s\geq20$. Based on this observation, the 20-stage configuration is a preferable setting to balance the reconstruction performance and computational costs. We fix the stage number $\ell_s=20$ for the natural image CS throughout all experiments.

\textbf{Test of geometric incremental domain number $n+1$:}
To explore the relationship between different geometric incremental domains and the reconstruction performance, we tune the geometric incremental domain number $n+1$ from 4 to 11. Fig.\ref{fig_net_parameters}(b) shows that the performances gradually improve when the geometric incremental spectral subspace number increases and fluctuates in a small range after 6. Taking into account the trade-off between network complexity and reconstruction performance, we set the geometric incremental domain number $n+1=6$ in all configurations.

\textbf{Test of filter number $N_{f}$:}
To explore the degree of parameterization over and under that influences reconstruction performance, we compare them in various number of filters. Fig.\ref{fig_net_parameters}(c) shows the average PSNR performance in Set11 for different filter numbers with CS ratio 25\%. When the filter number is $N_{f}=8$, the network works well even if it is under-parameterized and does not fully use the architecture's potential. We can find that even if the over-parameterization is severe, our method can still maintain excellent reconstruction performance without being troubled by over-fitting. Taking into account the trade-off between network complexity and reconstruction performance, we set the default filter number as 32 in all configurations.

\textbf{Test of with/without AR:}
To better explore the importance of AR, we analyze reconstruction results by varying the stage number $\ell_s$ from 10 to 30 at 5 intervals with/without AR. The average PSNR curves of the reconstructed results are shown in Fig.\ref{fig_net_parameters}(d). We find that the proposed AR can indeed enhance the reconstruction performance at different stages and works better on larger stages.

\textbf{Ablation study:}
Next, we conduct a group of ablation studies to evaluate the effectiveness of CI, AR, and N-II. Reconstruction performance comparisons are shown in Table \ref{table_ablation}. From variants (a) and (b), it is clear that the proposed CI can compensate for missing texture information from different geometric domains and improve reconstruction performance effectively.
Comparisons between variants (d) and (e) and between variants (c) and (f) further demonstrate the effectiveness of the proposed CI. Comparison between variants (a) and (c) shows that the proposed AR can achieve better reconstruction performance. Actually, AR is designed to approximate the remainder of operator spectral decomposition, and it is not surprising that its role is smaller than that of CI's. Compared to variant (a), variant (d) with the N-II scheme can improve reconstruction performance and provide a theoretical guarantee of faster convergence $(\mathcal{O}(\frac{1}{k^2}))$ than ISTA $(\mathcal{O}(\frac{1}{k}))$. The proposed CI, AR and N-II in our Nest-DGIL variant (g) can promote each other and achieve a great and stable improvement in reconstruction performance compared to baseline (a). These comparisons show that the CI is the most critical component.

\begin{table}
\centering
\setlength\tabcolsep{6pt}%
\caption{{Reconstruction performance comparisons with different combinations of CI, AR and N-II on Set11 with CS ratio 25\%.}}
\label{table_ablation}
    \begin{tabular}{cccccccc}
        \toprule[1.5pt]
\textbf{Variant} &(a) &(b) & (c) & (d)&{(e)}&{(f)}&(g)\\  \midrule[0.8pt]
\textbf{CI}&-&+&-&-&+&+&+\\
\textbf{AR} &-&-&+&-&-&+&+ \\
\textbf{N-II} &-&-&-&+&+&-&+\\
\textbf{PSNR}&{36.38}&{36.74}&{36.58}&{36.44}&{36.80}&{36.78}&{\textbf{36.81}}\\
        \bottomrule[1.5pt]
\end{tabular}
\end{table}

\begin{table}
\centering
\setlength\tabcolsep{10pt}%
\caption{Reconstruction performance comparisons (Average PSNR/Average SSIM) with different shared settings of proposed Nest-DGIL on Set11 with CS ratio 25\%.
}
\label{table_shared}
    \begin{tabular}{ccccc}
        \toprule[1.5pt]
        {\textbf{Variant}} &{\textbf{Shared setting}}& {\textbf{Parameters}} & {\textbf{PSNR/SSIM}}\\
\midrule[0.8pt]
        (a)  &Shared CI, AR &93099 &{36.37/0.9577}\\
        (b)  &Shared CI& 93175&{36.47/0.9585}\\
        (c)  & Shared AR &1861790 &{36.66/0.9599}\\
        (d)  &Unshared (default) &1861866&{\textbf{36.81/0.9603}}\\
        \bottomrule[1.5pt]
\end{tabular}
\end{table}

\textbf{Module sharing configurations:}
To demonstrate the flexibility of the proposed framework that does not have to be the same network parameter configurations in different stages, we conduct several variants of Nest-DGIL with different shared settings among stages. Table \ref{table_shared} lists the average PSNR and average SSIM comparisons for natural image CS with different shared settings of the proposed Nest-DGIL in Set11 with the CS ratio 25\%. Note that the best reconstruction performance is achieved when using the default unshared variant (d), which is the most flexible with the largest number of parameters. The variant (a) that shares both CI and AR in all stages is the least flexible with the smallest number of parameters and achieves the worst performance. If only CI or AR is shared, variants (b) and (c) increase average 0.10 dB and 0.29 dB over variant (a), respectively. Therefore, we adopt the default unshared variant (d) to perform the following experiments.

From the above comparisons, we can find that our method can also work well even if the network is under-parameterization so that it does not fully use the architecture's potential. In addition, while the network is overparameterized, our method can still maintain excellent reconstruction performance without obvious overfitting and degradation. Actually, we attribute the superiority of our method to two factors. Firstly, it has a cascade GIL module with geometric spectral decomposition and multi-regularisers truncation, which can obtain more texture compensation from different geometric decomposition domains. Secondly, the deep Nesterov-II scheme avoids the risk of intermediate reconstruction results falling outside the geometric domains.

\begin{table*}
    \setlength\tabcolsep{10pt}%
    \centering
    \newsavebox{\mybox}
    \begin{lrbox}{\mybox}
    {
    \begin{tabular}{clcccccc}
    \toprule[1.5pt]
     \multirow{3}{*}{\textbf{Dataset}} &\multirow{3}{*}{\textbf{Method}}&\multirow{3}{*}{\textbf{\shortstack{Sampling\\matrix}}}  & \multicolumn{5}{c}{\textbf{CS Ratio}} \\ \cmidrule(r){4-8}
        &&& 10\% & 25\% & 30\% & 40\% & 50\%  \\
         \midrule[0.8pt]
       \multirow{16}{*}{Set11}&DnCNN \cite{Zhang2017a}&\multirow{9}{*}{\shortstack{Fixed\\random\\Gaussian}}  &23.95/0.7040
 &26.92/0.8042&27.50/0.8270& 28.75/0.8539& 29.71/0.8768\\
       &ISTA-Net \cite{Zhang2018}&&23.42/0.6805&31.30/0.9066&32.86/0.9276&35.23/0.9504&37.36/0.9660\\
       &ISTA-Net+ \cite{Zhang2018}&& 26.41/0.8005&32.33/0.9206&33.73/0.9376&35.96/0.9564&38.03/0.9694 \\
      &DPDNN \cite{Dong2019}&& 24.75/0.7473&31.32/0.9093&32.75/0.9289&34.56/0.9457&36.82/0.9641\\
       &FISTA-Net \cite{Xiang2020} & &25.08/0.7545&30.26/0.8882&31.76/0.9124&33.80/0.9376&35.48/0.9521\\
       &AMP-Net-K \cite{Zhang2021}&& 26.22/0.7924&32.08/0.9172&33.55/0.9350&35.95/0.9566&37.97/0.9693\\
       &AMP-Net-K-B \cite{Zhang2021}&&27.78/0.8392&33.39/0.9340&34.68/0.9463&\underline{36.90}/\underline{0.9623}&\underline{38.72}/\underline{0.9724} \\
       &ISTA-Net++ \cite{You2021}&& \underline{28.23}/\underline{0.8480}&\underline{33.62}/\underline{0.9361}&\underline{34.80}/\underline{0.9474}& 36.84/0.9622&38.62/0.9722\\
      &{\textbf{${\text{Nest-DGIL}}$ }}& &{\textbf{30.35}/\textbf{0.8875}}& {\textbf{36.81}/\textbf{0.9603}}&       {\textbf{38.26}/\textbf{0.9692}}&{\textbf{40.83}/\textbf{0.9806}}&{\textbf{43.24}/\textbf{0.9876}}\\\cmidrule(r){2-8}
      &COAST \cite{You2021a}&\multirow{7}{*}{\shortstack{Jointly\\learned}}& 28.48/0.8584&33.88/0.9396&35.05/0.9496&37.20/0.9645&39.01/0.9742\\ &OPINE-Net\cite{Zhang2020}& & {29.64}/0.8882&   34.73/0.9516&36.16/0.9609&  38.45/0.9732&   40.47/0.9811\\
     &AMP-Net-K-M\cite{Zhang2021} && 29.27/0.8853&34.56/0.9517& 36.05/0.9620&   38.35/0.9741&   {40.56}/0.9822\\
      &AMP-Net-K-BM\cite{Zhang2021}& &{29.73}/{0.8934}& {34.87}/{0.9545}&   \underline{36.35}/\underline{0.9641}&\underline{38.62}/\underline{0.9753}&40.55/{0.9826}  \\
      & {$\text{CSformer}_{bsd400}$ \cite{Ye2023}}&&{{29.79}/0.8883}& {34.81/0.9527}&{---/---}&{---/---}&{40.73/0.9824}\\
      & {$\text{CSformer}_{coco}$ \cite{Ye2023}}&&{\textbf{30.66}/\underline{0.9027}}&{ \underline{35.46}/\underline{0.9570}}&{---/---}&{---/---}&{\underline{41.04}/\underline{0.9831}}\\
     &{\textbf{${\text{Nest-DGIL+}}$ }} &&{\underline{30.44}/\textbf{0.9038}}&      {\textbf{36.37}/\textbf{0.9626}}&{\textbf{37.77}/\textbf{0.9702}}&{\textbf{40.25}/\textbf{0.9805}}&{\textbf{42.30}/\textbf{0.9866}}\\
      \midrule[0.8pt]
       \multirow{16}{*}{BSD68}&DnCNN \cite{Zhang2017a}&\multirow{9}{*}{\shortstack{Fixed\\random\\Gaussian}}  &24.24/0.6506&26.61/0.7602&27.14/0.7876&28.21/0.8257&29.15/0.8547\\
       &ISTA-Net \cite{Zhang2018}&& 23.82/0.6298&28.95/0.8408&30.12/0.8727&32.09/0.9135&34.06/0.9426\\
       &ISTA-Net+ \cite{Zhang2018}&& 25.42/0.7031&29.37/0.8513&30.51/0.8810&32.38/0.9185&34.36/0.9455 \\
      &DPDNN \cite{Dong2019}&&24.70/0.6748&29.09/0.8476&30.23/0.8770&31.80/0.9099& 33.84/0.9410\\
       &FISTA-Net \cite{Xiang2020} && 24.89/0.6796&28.50/0.8286&29.59/0.8609    &31.26/0.8990&32.85/0.9262\\
       &AMP-Net-K \cite{Zhang2021}&&25.37/0.7028&29.34/0.8510&30.46/0.8797& 32.44/0.9190&34.36/0.9454\\
       &AMP-Net-K-B \cite{Zhang2021}&& {26.12}/0.7310&30.02/0.8667&31.11/0.8925&33.03/\underline{0.9272}&\underline{34.92}/\underline{0.9508}\\
       &ISTA-Net++ \cite{You2021}&& \underline{26.28}/\underline{0.7377}&\underline{30.15}/\underline{0.8690}&\underline{31.18}/\underline{0.8931}&\underline{33.04}/0.9269&34.84/0.9499\\
      &{\textbf{${\text{Nest-DGIL}}$ }} &&{\textbf{28.08}/\textbf{0.7868}}&     {\textbf{32.96}/\textbf{0.9143}}&{\textbf{34.34}/\textbf{0.9355}}&{\textbf{36.81}/\textbf{0.9615}}&{\textbf{39.43}/\textbf{0.9784}}
      \\\cmidrule(r){2-8}
      &COAST \cite{You2021a}&\multirow{7}{*}{\shortstack{Jointly\\learned}}& 26.40/0.7441&30.30/0.8735&31.30/0.8964&33.23/0.9301&35.09/0.9526\\
      &OPINE-Net\cite{Zhang2020}&& {27.86}/0.8075&31.64/0.9095&32.80/0.9290&34.85/0.9540&36.83/0.9700\\
     &AMP-Net-K-M\cite{Zhang2021} && 27.71/0.8061&31.67/0.9130  &32.79/0.9306&34.89/{0.9556}&{36.97}/{0.9715}\\
      &AMP-Net-K-BM\cite{Zhang2021}&& {27.98}/\underline{0.8143}&\underline{31.91}/\underline{0.9170}&\underline{33.04}/\underline{0.9345}&\underline{35.09}/\underline{0.9577}&{37.08}/{0.9728} \\
      & {$\text{CSformer}_{bsd400}$ \cite{Ye2023}}&&{{28.05}/0.8045}&{31.82/0.9106}&{---/---}&{---/---}&{{37.14}/\underline{0.9766}}\\
      & {$\text{CSformer}_{coco}$ \cite{Ye2023}}&&{\underline{28.28}/0.8078}&{\underline{31.91}/0.9102}&{---/---}&{---/---}&{\underline{37.16}/0.9714}\\
      &{\textbf{${\text{Nest-DGIL+}}$ }} &&{\textbf{28.64}/\textbf{0.8274}}&{\textbf{33.11}/\textbf{0.9298}}&       {\textbf{34.26}/\textbf{0.9451}}&{\textbf{36.56}/\textbf{0.9667}}&{\textbf{38.77}/\textbf{0.9797}}\\
      \midrule[0.8pt]
       \multirow{16}{*}{Urban100}&DnCNN \cite{Zhang2017a}&\multirow{9}{*}{\shortstack{Fixed\\random\\Gaussian}}  & 21.81/0.6285&24.45/0.7494&25.04/0.7747&26.21/0.8110&27.25/0.8453\\
       &ISTA-Net \cite{Zhang2018}&& 21.28/0.6057&27.99/0.8601&29.52/0.8905&32.03/0.9301&34.39/0.9554\\
       &ISTA-Net+ \cite{Zhang2018}&& 23.60/0.7190&29.12/0.8836&30.50/0.9090&    32.86/0.9404&35.07/0.9600 \\
      &DPDNN \cite{Dong2019}&& 22.52/0.6722&28.14/0.8669&29.67/0.8952&31.63/0.9260&33.85/0.9523\\
       &FISTA-Net \cite{Xiang2020} && 22.51/0.6714& 26.81/0.8329&28.22/0.8677&30.10/0.9055&31.81/0.9296\\
       &AMP-Net-K \cite{Zhang2021}&& 23.41/0.7120&28.82/0.8773&30.33/0.9040&    32.77/0.9382&34.92/0.9589\\
       &AMP-Net-K-B \cite{Zhang2021}&& 24.85/0.7640&30.23/0.9027&31.63/0.9242&\underline{33.94}/0.9496&\underline{35.97}/0.9644\\
       &ISTA-Net++ \cite{You2021}&& \underline{25.29}/\underline{0.7772}&\underline{30.63}/\underline{0.9089}&\underline{31.86}/\underline{0.9269}&33.93/\underline{0.9502}&35.79/\underline{0.9651}\\
      &{\textbf{${\text{Nest-DGIL}}$ } } &&{\textbf{28.97}/\textbf{0.8502}}&        {\textbf{35.47}/\textbf{0.9520}}&{\textbf{37.08}/\textbf{0.9647}}&{\textbf{39.78}/\textbf{0.9787}}&{\textbf{42.36}/\textbf{0.9876}}\\
      \cmidrule(r){2-8}
      &COAST \cite{You2021a}&\multirow{7}{*}{\shortstack{Jointly\\learned}}&25.67/0.7979
      &31.04/0.9161&32.25/0.9319&34.38/0.9537&36.24/0.9675\\
      &OPINE-Net\cite{Zhang2020}&& {26.55}/0.8343&{31.47}/0.9289&\underline{32.79}/0.9441&\underline{34.95}/0.9627&{36.87}/0.9743\\
     &AMP-Net-K-M\cite{Zhang2021} && 25.97/0.8245&30.96/0.9266&32.21/0.9410&    34.54/0.9617&36.65/0.9744\\
      &AMP-Net-K-BM\cite{Zhang2021}&& 26.37/{0.8377}&31.22/{0.9309}&32.52/\underline{0.9449}&34.67/\underline{0.9631}&36.53/{0.9744}  \\
      & {$\text{CSformer}_{bsd400}$ \cite{Ye2023}}&&{{27.92}/0.8458}&{{32.43}/0.9332}&{---/---}&{---/---}&{{37.88}/0.9766}\\
      & {$\text{CSformer}_{coco}$ \cite{Ye2023}}&&{\textbf{29.61}/\textbf{0.8762}}&{\underline{34.16}/\underline{0.9470}}&{---/---}&{---/---}&{\underline{39.46}/\underline{0.9811}} \\
      &{\textbf{${\text{Nest-DGIL+}}$ }} &&{\underline{28.81}/\underline{0.8664}}&{\textbf{34.51}/\textbf{0.9516}}&{\textbf{35.75}/\textbf{0.9619}}&{\textbf{38.07}/\textbf{0.9762}}&{\textbf{40.14}/\textbf{0.9846}}\\
    \bottomrule[1.5pt]
\end{tabular}}
\end{lrbox}
    \caption{{Reconstruction performance comparisons (Average PSNR/Average SSIM) for natural image CS with different CS ratios. The best and second best results are highlighted in bold font and underlined ones, respectively.}}
    \scalebox{1}{\usebox{\mybox}}
    \label{table_result_NI}
\end{table*}

\begin{figure*}[htbp]
\centering
\includegraphics[width=1\textwidth]{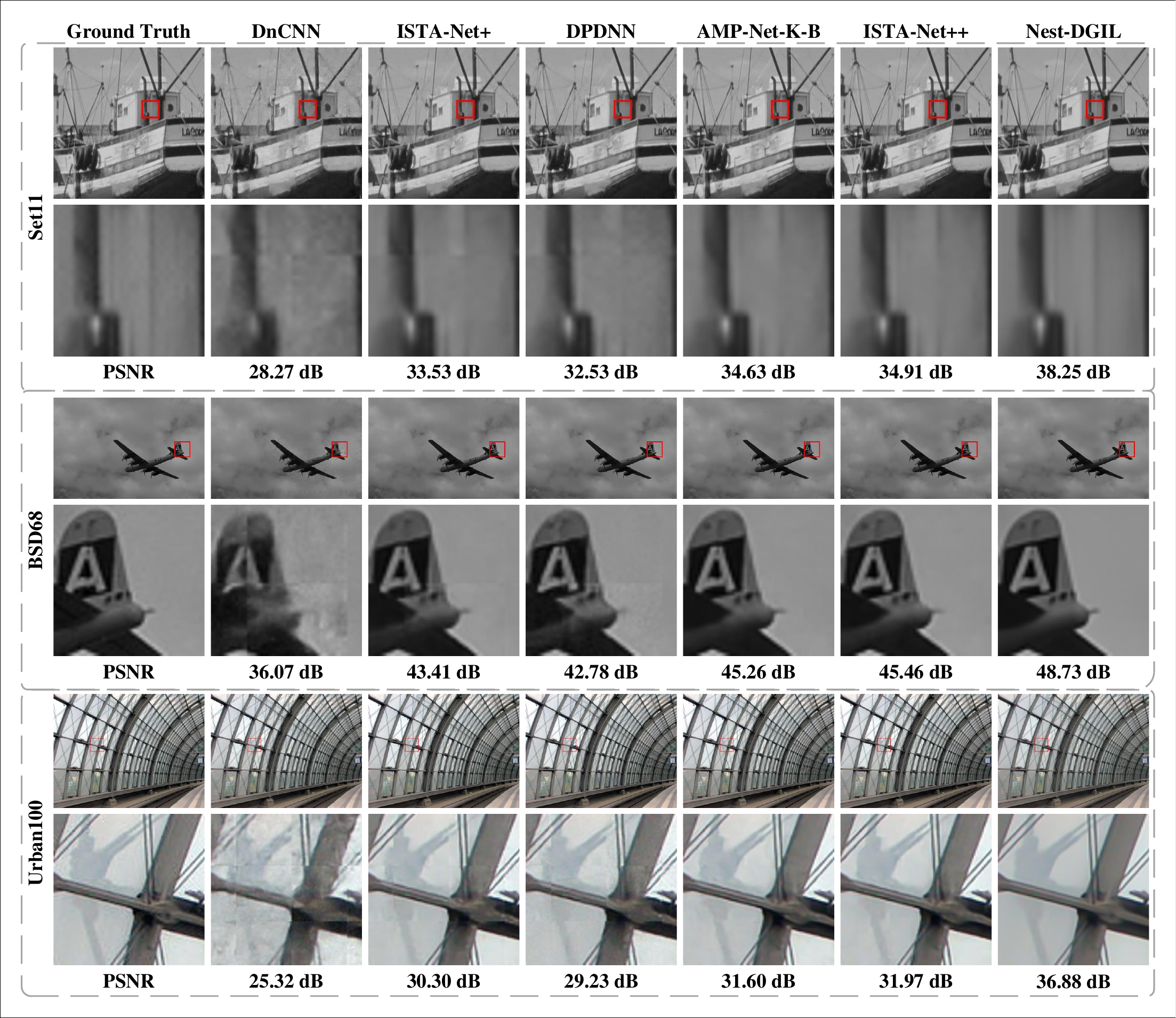}
\caption{{Reconstruction comparisons on Set11, BSD68 and Urban100 with CS ratio 25\% by different methods with fixed random Gaussian matrix.}}
\label{fig_comparison_image}
\end{figure*}

\subsection{Comparisons with other state-of-the-art methods}
To demonstrate the superiority of our framework, we compare it with other widely used methods on two CS tasks (natural image CS and sparse-view CT).

\textbf{Natural image CS:}
We compare our Nest-DGIL and Nest-DGIL+ with several recent representative CS reconstruction methods, including the deep learning method (DnCNN \cite{Zhang2017a}, CSformer \cite{Ye2023}) and state-of-the-art deep unfolding methods (ISTA-Net \cite{Zhang2018}, ISTA-Net+ \cite{Zhang2018}, DPDNN \cite{Dong2019}, FISTA-Net \cite{Xiang2020}, AMP-Net-K \cite{Zhang2021}, AMP-Net-K-B \cite{Zhang2021}, ISTA-Net++ \cite{You2021}, COAST \cite{You2021a}, OPINE-Net \cite{Zhang2020}, AMP-Net-K-M \cite{Zhang2021}, AMP-Net-K-BM \cite{Zhang2021}) to demonstrate our method's excellent performance on natural image CS reconstruction. Following \cite{Zhang2018}, the stage number of FISTA-Net is configured as 9. The results of CSformer \cite{Ye2023} are obtained by their public pretrained model. All the other compared methods are trained with the same training dataset \cite{Chen2017a} and the same methods with the corresponding works.

The average PSNR/SSIM values of the natural image CS reconstruction corresponding to five CS ratios with fixed random Gaussian and jointly learned sampling matrix are shown in Table \ref{table_result_NI}. For AMP-Net \cite{Zhang2021}, the variant AMP-Net-K-B with deblocking modules can effectively improve reconstruction results than the variant AMP-Net-K that only has denoising modules. Due to the cross-block strategy in the sampling process, ISTA-Net++ achieves state-of-the-art results. Our method almost outperforms other compared methods in all CS ratios obviously, and achieves average 1.58 dB, 1.37 dB and 2.54 dB improvement on Set11, BSD68 and Urban100. The reconstructed pixels on an inner narrow band of the patch boundary lack context information and usually have a serious block effect. This local block effect, although a small proportion of the overall, will greatly reduce PSNR. Performing the weighted reconstruction of the overlapping patches can reduce the block effect part of the edge reconstruction and significantly improves the PSNR results \cite{Ye2023}.

\begin{table*}[htbp]
    \caption{{Reconstruction performance comparisons (Average PSNR/Average RMSE(HU)) with different down-sampled projections for sparse-view CT. The best and second best results are highlighted in Bold font and underlined ones, respectively.}}
    \label{table_result_CT}
    \setlength\tabcolsep{7.5pt}%
    \centering
    \begin{tabular}{lcccccccc}
    \toprule[1.5pt]
         \multirow{3}{*}{\textbf{Method}} & \multicolumn{4}{c}{\textbf{LDCT-Data}\cite{Moen2020}} & \multicolumn{4}{c}{{\textbf{FRPLung-Data}\cite{Rusu2017}}} \\ \cmidrule(r){2-5} \cmidrule(r){6-9}
         &  60 views & 90 views& 120 views& 180 views &  60 views& 90 views& 120 views& 180 views\\
         \midrule[0.8pt]
        FBP &28.69/151.3&31.98/103.7&34.67/76.06 &39.03/46.08&28.88/147.9&32.10/102.0&  34.70/75.57&    38.84/46.94\\
        {FISTA-TV} \cite{Beck2009a}&35.17/71.60&38.13/50.87 &40.24/39.90 &43.56/27.24 &35.06/72.83  &   38.12/51.16 &   40.25/39.94     &43.50/27.53 \\
        {RED-CNN} \cite{Chen2017} & 42.01/32.65&43.45/27.69&44.44/24.78&45.80/21.18&40.88/37.11 &   42.89/29.45  &  44.21/25.31  &45.97/20.68 \\
        {FBPConvNet} \cite{Jin2017}& 42.26/31.74  &43.50/27.57 &44.44/24.78 & 45.97/20.81 &41.16/35.88 &    42.98/29.16 &44.22/25.27 &46.05/20.48 \\
        {DU-GAN} \cite{Huang2022} &38.70/47.72 &39.91/41.53 &41.36/35.18 &43.20/28.51& 37.85/52.51  &   39.48/43.54  & 41.21/35.72 &43.11/28.67 \\
        Deep Decoder \cite{Heckel2019} &{34.85/74.67}&{37.90/52.39}& {38.79/47.35}&{39.24/44.93}& {34.94/73.65}&    {37.87/52.47}&{38.95/46.45}  &{39.40/44.15} \\
        {PD-Net}\cite{Adler2018} & 41.22/35.76  &42.26/31.74 &42.43/31.13 &43.85/26.38 &39.47/43.62  &  40.43/39.12  &  40.85/37.27  &41.67/33.96 \\
        ISTA-Net\cite{Zhang2018}&43.38/27.98 &44.40/24.94&45.17/22.86 &46.52/19.58 &43.16/28.51 &44.63/24.13  &45.59/21.63  &47.07/18.27 \\
        ISTA-Net+ \cite{Zhang2018} &\underline{43.52}/\underline{27.57} &\underline{44.50}/\underline{24.70}&\underline{45.23}/\underline{22.69} & \underline{46.60}/\underline{19.42} &\underline{43.36}/\underline{27.89} &   \underline{44.77}/\underline{23.72} &\underline{45.65}/\underline{21.46} &\underline{47.11}/\underline{18.19}\\
       FISTA-Net \cite{Xiang2020} &37.31/56.16  &39.28/44.77 &40.68/38.13&41.65/34.08 & 36.80/59.35     &39.04/45.88 &40.53/38.63 &41.60/34.12\\
       AMP-Net-K \cite{Zhang2021} &38.64/48.09 &43.55/27.44& 42.72/30.19 &44.83/23.72 & 37.94/52.06 &   43.77/26.58 &42.77/29.82&45.06/22.94 \\
       Nest-DGIL &  \textbf{43.77}/\textbf{26.79}  & \textbf{44.67}/\textbf{24.21}  & \textbf{45.38}/\textbf{22.32}  & \textbf{46.72}/\textbf{19.17} & \textbf{43.77}/\textbf{26.62}     &\textbf{44.93}/\textbf{23.31}& \textbf{45.76}/\textbf{21.22} & \textbf{47.16}/\textbf{18.06}\\
       \bottomrule[1.5pt]
    \end{tabular}
\end{table*}

\begin{figure*}[htbp]
\centering
\includegraphics[width=1\textwidth]{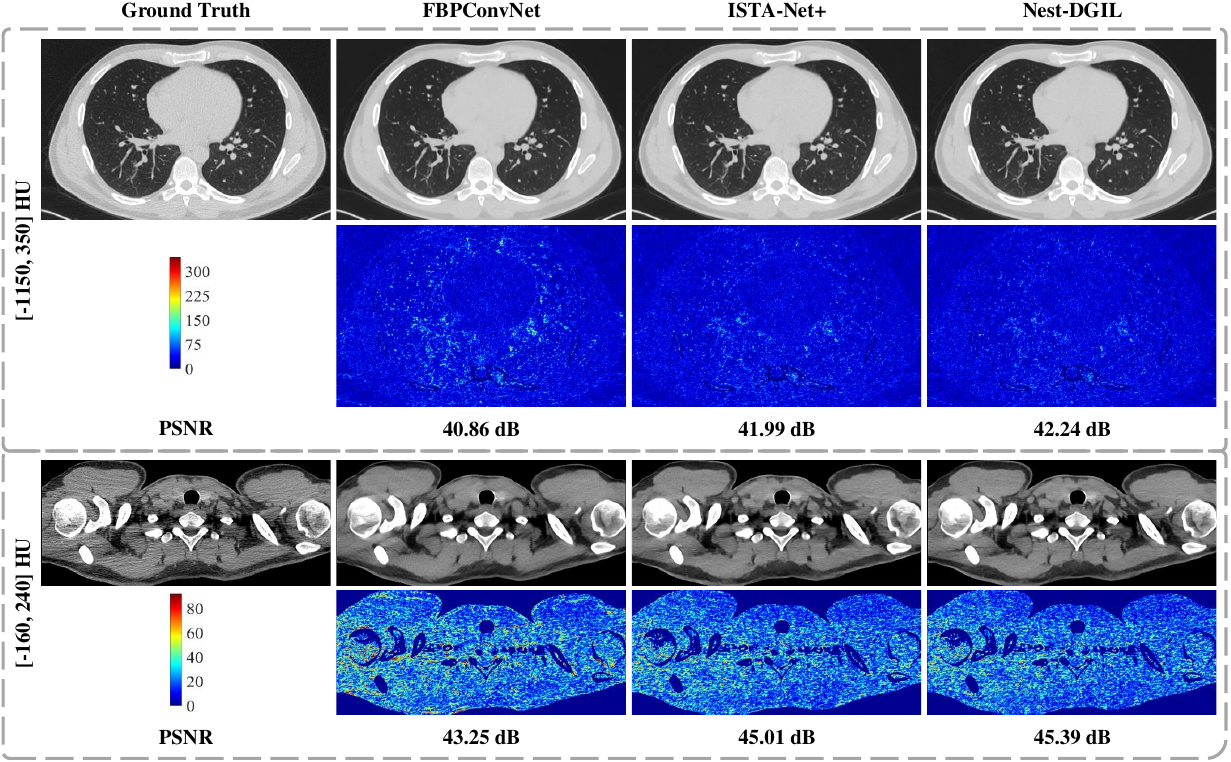}
\caption{The axial reconstruction results from different methods for parallel beam projection with 60 views on LDCT-Data.}
\label{fig_comparison_CT}
\end{figure*}

Fig.\ref{fig_comparison_image} shows the visualizations of the reconstructed results for natural images with the CS ratio 25\% when comparing the proposed Nest-DGIL framework with state-of-the-art methods, where visual comparisons consist of full images and zoom-in details. We can observe that other compared methods are obviously block-effect, and details are not recovered very well or even lost. Our method can effectively reduce block-effect (\emph{e.g.} chimney, cloud and glass in zoom-in details) and restore much more texture details, yielding much better visually pleasant results.

\textbf{Sparse-view CT:}
To further demonstrate the superiority of our approach, we perform a group of comparisons to evaluate the reconstruction performance in sparse-view CT. We compare our method with classical methods (FBP, FISTA-TV \cite{Beck2009a}), deep learning method (RED-CNN \cite{Chen2017}, FBPConvNet \cite{Jin2017}, DU-GAN \cite{Huang2022}, Deep Decoder \cite{Heckel2019}) and state-of-the-art deep unfolding methods (PD-Net \cite{Adler2018}, ISTA-Net \cite{Zhang2018}, ISTA-Net+ \cite{Zhang2018}, FISTA-Net \cite{Xiang2020}, AMP-Net-K \cite{Zhang2021}). The FBP with the "Ram-Lak" filter, which is implemented with the iradon transform in TorchRadon \cite{Ronchetti2020}, is adopted to provide initialization for the proposed method and other compared deep models. The maximum number of iterations of FISTA-TV is set as 100 and the regularization parameter is tuned to optimal. Following \cite{Xiang2020}, the stage numbers of PD-Net, ISTA-Net, ISTA-Net +, FISTA-Net, AMP-Net-K, and Nest-DGIL are configured as 7. The number of iterations of the deep decoder is set to 2000 \cite{Heckel2019}.

\begin{table}
    \caption{The learned parameters at different stages with CS Ratio 25\%, and reconstruction performance (average PSNR) on Set11.}
    \label{table-NI-iteration}
    \setlength\tabcolsep{2pt}
    \centering
    \begin{tabular*}{\hsize}{@{}@{\extracolsep{\fill}}ccccccccc@{}}
    \toprule[1.5pt]
        Stage$(k)$ & $\mu_{k}$ & $\theta_{\tau_k}$ & $\gamma_{k}$& $\beta_{k,1}$ & $\beta_{k,2}$ & $\beta_{k,3}$ & $\beta_{k,4} $ &PSNR\\ \midrule[0.8pt]
        0&-&-&-&-&-&-&-&7.64\\
        1   & 2.380&1.20$\times10^{-1}$&0.595&1.086 &0.999 &0.931&0.887&{13.66}\\
        2   & 2.342&6.06$\times10^{-2}$&0.721&1.166&0.965&0.860&0.847&{21.27}\\
        3   & 2.304 &3.01$\times10^{-2}$&0.836&1.216&0.790&0.786&0.791&{26.31}\\
        4   & 2.266&1.49$\times10^{-2}$&0.917&1.255&0.800&0.765&0.760&{27.05}\\
        5   & 2.229&7.31$\times10^{-3}$&0.963&0.984&1.023&1.015&1.012&{27.16}\\
        6   & 2.191&3.59$\times10^{-3}$&0.985&1.048 &0.951&0.948&0.945 &{28.53}\\
        7   & 2.154&1.76$\times10^{-3}$&0.994&0.972&1.032&1.029&1.027&{28.97}\\
        8   & 2.117&8.62$\times10^{-4}$&0.998&0.859&1.141&1.141&1.141&{29.33}\\
        9   & 2.080&4.22$\times10^{-4}$&0.999&  0.949&1.052&1.052&1.052&{30.67}\\
        10   & 2.044&2.07$\times10^{-4}$&1.000&0.840&1.153&1.153&1.154&{31.05}\\
        11   & 2.007&1.01$\times10^{-4}$&1.000&0.923&1.078&1.079&1.079&{31.93}\\
        12   & 1.971&4.97$\times10^{-5}$&1.000&0.926&1.077&1.075&1.076&{29.89}\\
        13   & 1.935&2.43$\times10^{-5}$&1.000&0.824&1.159&1.159&1.159&{32.98}\\
        14   & 1.899&1.19$\times10^{-5}$&1.000&0.838&1.140&1.140&1.140&{34.25}\\
        15   & 1.864&5.84$\times10^{-6}$&1.000&1.084&0.927&0.927&0.927&{34.67}\\
        16   & 1.829&2.86$\times10^{-6}$&1.000&0.918&1.083&1.083&1.083&{34.80}\\
        17   & 1.793&1.40$\times10^{-6}$&1.000&0.576&1.305&1.305&1.305&{35.32}\\
        18   & 1.759&6.86$\times10^{-7}$&1.000&0.482&1.315&1.315&1.315&{35.69}\\
        19   & 1.724&3.36$\times10^{-7}$&1.000&0.702&1.198&1.198&1.198&{36.04}\\
        20   & 1.690&1.65$\times10^{-7}$&1.000&0.809&1.123&1.123&1.123&{36.81}\\
         \bottomrule[1.5pt]
    \end{tabular*}
\end{table}

\begin{table}
    \caption{The learned parameters at different stages with 60 views, and reconstruction performance (average PSNR) on {LDCT-Data}.}
    \label{table-CT-iteration}
    \renewcommand\arraystretch{1.1}
    \setlength\tabcolsep{2pt}
    \centering
    \begin{tabular*}{\hsize}{@{}@{\extracolsep{\fill}}ccccccccc@{}}
    \toprule[1.5pt]
        Stage$(k)$ & $\mu_{k}$ & $\theta_{\tau_k}$ & $\gamma_{k}$& $\beta_{k,1}$ & $\beta_{k,2}$ & $\beta_{k,3}$ & $\beta_{k,4} $ &PSNR\\ \midrule[0.8pt]
        0&-&-&-&-&-&-&-&28.69\\
        1   & 1.052&0.1027&0.595&0.979&0.975&1.019&1.025&20.31\\
        2   & 0.964&0.0415&0.721&0.734&1.965&0.915 &0.836&26.50\\
        3   & 0.880&0.0165&0.836&0.462&2.363&1.176&1.068&30.45\\
        4   & 0.801&0.0065&0.917&0.482&1.809&1.310&1.253&33.25\\
        5   & 0.726&0.0025&0.963&0.250&2.106&1.570&1.522&39.12\\
        6   & 0.657&0.0010&0.985&0.087&2.606&1.423&1.386 &41.71\\
        7   & 0.592&0.0004&0.994&0.192&1.923&1.642&1.628 &43.77\\ \bottomrule[1.5pt]
    \end{tabular*}
\end{table}

\begin{figure*}[htbp]
    \centering
    \includegraphics[width=0.98\textwidth]{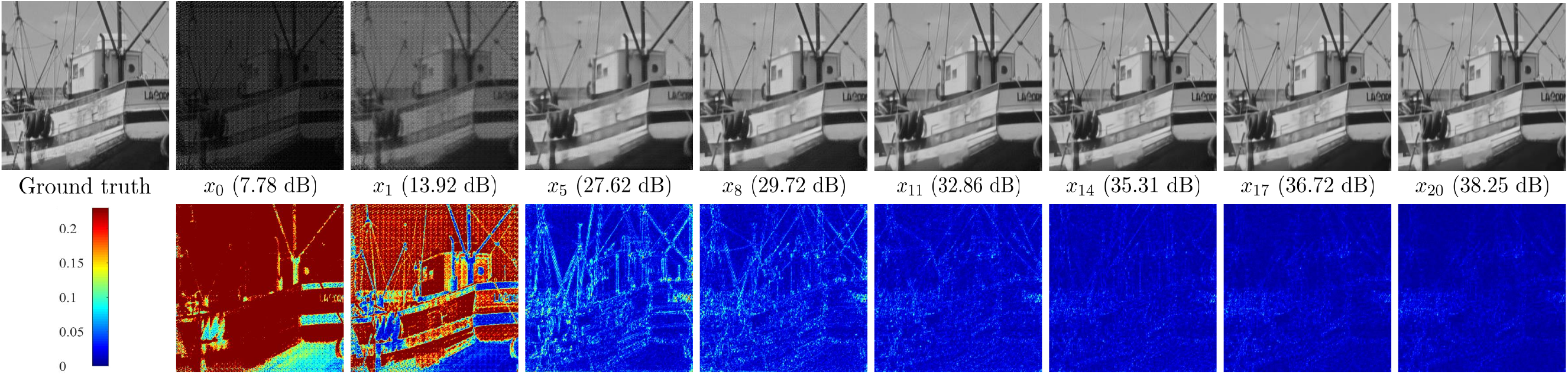}
    \caption{{Reconstructed intermediate boats images and the corresponding residual from CS ratio 25\% by Nest-DGIL at different stages.}}
    \label{fig_stage_intetimate}
\end{figure*}

Table \ref{table_result_CT} lists the average PSNR and RMSE (HU) of the compared methods with different down-sampled projection views. We can observe that ISTA-Net and ISTA-Net+ consistently outperform FISTA-Net for all downsampled projections due to unshared learnable parameters. Due to abundant learnable parameters and considerable training data, RED-CNN and FBPConvNet achieve a good reconstruction performance. Due to underparameterization and without checking the data consistency with the measurement, the Deep Decoder cannot reconstruct the CT image well. Our method outperforms comparison methods at all down-sampled projections and achieves an average 0.17 dB improvement on LDCT-Data and an average 0.18 dB improvement on FRPLung-Data. In addition, our method can obtain a more accurate HU reconstruction and provide a better service for clinician diagnosis.

The results of axial reconstruction from different methods for parallel beam projection with 60 views are shown in Fig.\ref{fig_comparison_CT}. FBPConvNet can remove streak artifacts effectively, but some tiny structures could be smoothed out. Although ISTA-Net+ can remove some noise and streaking artifacts, but results in incomplete preservation of details and texture information. Due to cascade geometric incremental learning and adaptive remainder optimization, our method achieves the best reconstruction performance in terms of noise artifact removal and detail preservation (e.g., tiny blood vessels and bronchi).

\subsection{Analysis for intermediate results}
As introduced in previous section, the proposed framework is an iterative architecture, so it is interesting and meaningful to evaluate the intermediate results and the learned parameters in different stages. The learned parameters for natural image CS with CS ratio 25\% and sparse view CT with 60 views at different stages are shown in Table \ref{table-NI-iteration} and Table \ref{table-CT-iteration}. The reconstruction performances become better with the increasing of iterations. We can see that the parameters $\mu_{k}$ and $\theta_{\tau_k}$ decrease monotonically, while $\gamma_{k}$ increases with respect to $k$. The learned parameters $\mu_{k}$, $\theta_{\tau_k}$ and $\gamma_{k}$ are consistent with the parameter configuration in traditional model-based reconstruction. It implies that our approach fully inherits the characteristics and advantages of model-based methods. For more explanation, we show the images of intermediate boats reconstructed and the corresponding residual from the CS ratio 25\% by Nest-DGIL at different stages in Fig.\ref{fig_stage_intetimate}. We can observe that block-effect removal and detail recovery are performed gradually across the stages. The trained end-to-end Nest-DGIL with a meaningful model-based network architecture not only facilitates the enhancement of intermediate image results but also contributes to final better performance by supervising a smoothing reconstruction flow.

In addition, we can find that the weight coefficients $\beta_{k,1}$, $\beta_{k,2}$, $\beta_{k,3}$ and $\beta_{k,4}$ work at different stages. The results demonstrate that the proposed adaptive spectral decomposition residual part $\epsilon_{k,n}$ mainly restore the missing texture information truncated by the principal part of the spectral geometric incremental decomposition. Therefore, adaptive initialization, adaptive spectral decomposition remainder and learnable parameters setting are of great importance for model flexibility and ensuring converging smoothly.

\begin{table}[htbp]
    \caption{{Reconstruction performance comparisons (Average PSNR) on different training datasets.}}
    \label{table_training_set}
    \renewcommand\arraystretch{1.1}
    \setlength\tabcolsep{5pt}
    \centering
    \begin{tabular}{llccccc}
    \toprule[1.5pt]
        \multirow{3}{*}{\textbf{Dataset}} & \multirow{3}{*}{\textbf{Method}} & \multicolumn{4}{c}{\textbf{CS Ratio}} \\ \cmidrule(r){3-7}
        & &  10\% & 25\% & 30\% & 40\% & 50\%\\
         \midrule[0.8pt]
        \multirow{2}{*}{\textbf{BSD68}}&${\text{Nest-DGIL+}}^{\text{Set11}}$&28.10&32.12&33.23&35.34&37.30 \\
        &Nest-DGIL+&28.64&33.11&34.26&36.56&38.77\\
        \multirow{2}{*}{\textbf{Urban100}}&${\text{Nest-DGIL+}}^{\text{Set11}}$&27.53&32.22&33.45&35.69&37.55\\
        &Nest-DGIL+&28.81&34.51&35.75&38.07&40.14\\
       \bottomrule[1.5pt]
    \end{tabular}
\end{table}

\subsection{Generalizability evaluation on different training datasets}
For natural image CS tasks, Nest-DGIL+ is trained based on 400 training images \cite{Chen2017a}. To further evaluate the generalizability of our model, we train it in Set11 with only 11 images and evaluate it in BSD68 and Urban100. The results are shown in Table \ref{table_training_set}. As a result of the fact that Set11 is very small, the potential of ${\text{Nest-DGIL+}}^{\text{Set11}}$ is not fully released. The reconstruction performances of ${\text{Nest-DGIL+}}^{\text{Set11}}$ drop 1.05 dB on BSD68 and 2.17 dB on Urban100, but are still excellent. It demonstrates the excellent generalizability of our method.

\section{Conclusions}\label{secIV}
In this study, we propose a Nesterov-informed geometric incremental learning framework (Nest-DGIL) based on the second Nestirov proximal gradient optimization. It not only has the powerful learning ability for high/low frequency image features but also can theoretically guarantee that more geometric texture details will be reconstructed from preliminary linear reconstruction. Our Nest-DGIL network can avoid the risk of intermediate reconstruction results falling outside the geometric domain and achieve fast convergence. Such a Nesterov-informed learnable architecture gives us a new perspective to design explainable networks. Extensive experiments show that the proposed Nest-DGIL framework can greatly improve reconstruction performance on the existing state-of-the-art methods in different applications (natural image CS and sparse-view CT). Our architecture has good generalizability for image reconstruction due to the proximal gradient-based optimization unfolding and cascade incremental learning and can be potentially applied to other inverse problems in imaging.

{\appendices
\section{Lemma 1 and Lemma 2} \label{appendix-C}

\begin{lemma}[\cite{Beck2009a}]
Let $\left\{\bm{x}_{k}\right\}$ be the sequence generated by the proximal gradient method ((\ref{eq4-2}) to (\ref{eq5-2})). Therefore, the reconstruction algorithm based on proximal gradients satisfies the linear rate of convergence $(\mathcal{O}(\frac{1}{k}))$, i.e.
\[\mathcal{F}\left(\bm{x}_{k}\right)-\mathcal{F}\left(\bm{x}^{*}\right) \leq \frac{\overline{\mathcal{L}}\left\|\bm{x}_{0}-\bm{x}^{*}\right\|^{2}}{2 k}\]
for some constant $\overline{\mathcal{L}}$ and any optimal solution $\bm{x}^{*}$.
\label{lemma1}
\end{lemma}

\begin{lemma}[\cite{ANSELONE197467}]
 Let $\mathcal{S}$ be the Banach space of functions $\bm{x}(t)$, $a \leqslant t \leqslant b$, with the uniform norm, $\|\bm{x}\|=\max |\bm{x}(t)|$. Let $\mathcal{K}$ be an integral operator on $\mathcal{S}$, i.e.
\[(\mathcal{K}\bm{x})(s)=\int_{0}^{1} \bm{k}(s, t) \bm{x}(t) d t, \quad a \leqslant t \leqslant b\]
with a continuous kernel $\bm{k}$. Hence one has
$$
(I-\epsilon \mathcal{K})^{-1}=\sum_{n=0}^{\infty} \epsilon^{n} \mathcal{K}^{n},
$$
where $0<\epsilon<1 /\bm{k}_{\text{max}}$, where $\bm{k}_{\text{max}}=\max\; \bm{k}(s, t)$, and the series converges in the operator norm.
\label{lemma2}
\end{lemma}

\section{More explanation about the left- and right-hand operators for odd and even numbers} \label{appendix-A}
As we have defined $\mathcal{M}(\bm{x})=\mathcal{K}^{\prime}\left(\psi\left(\mathcal{K}(\bm{x})\right)\right)$ and $\bm{\omega}_{k,i}=\mathcal{M}^{i}(\bm{m}_{k})$ ($i=1,\dots,n$), we have $$\begin{aligned}
        \bm{\omega}_{k,2}&=\underbrace{\underline{\mathcal{K}^{\prime}\psi(\mathcal{K}}[\underline{\mathcal{K}^{\prime}\psi(\mathcal{K}}}_{2}(\bm{m}_{k}))]),\\ \bm{\omega}_{k,3}&=\underbrace{\underline{\mathcal{K}^{\prime}\psi(\mathcal{K}}[\underline{\mathcal{K}^{\prime}\psi(\mathcal{K}}[\underline{\mathcal{K}^{\prime}\psi(\mathcal{K}}}_{3}(\bm{m}_{k}))]).
    \end{aligned}$$

    In fact, the operator $\mathcal{K}^{\prime}$ is the transpose of the feature extractor $\mathcal{K}$ and can be understood as the inverse operator of $\mathcal{K}$. We can further separate operators $\mathcal{K}^{\prime}(\cdot)$ and $\psi\left(\mathcal{K}(\cdot)\right)$ as follows
    $$\begin{aligned}
        \bm{\omega}_{k,2}&=\underbrace{\underline{\mathcal{K}^{\prime}}\underline{\psi(\mathcal{K}}}_{right}\underbrace{[\underline{\mathcal{K}^{\prime}}\underline{\psi(\mathcal{K}}}_{left}(\bm{m}_{k}))])],\\ \bm{\omega}_{k,3}&=\underbrace{\underline{\mathcal{K}^{\prime}}\underline{\psi(\mathcal{K}}[\underline{\mathcal{K}^{\prime}}}_{right}\underbrace{\underline{\psi(\mathcal{K}}[\underline{\mathcal{K}^{\prime}}\underline{\psi(\mathcal{K}}}_{left}(\bm{m}_{k}))])],
    \end{aligned}$$
    where left- and right-hand operators are the total transform feature extractor and the inverse transform feature extractor.

    When $\bm{\omega}_{k,2}$ and $\bm{\omega}_{k,3}$ are the truncation remainders, we can embed the shrinkage-thresholding operator as follows $$\begin{aligned}
        \bm{\omega}_{k,2}&=\underline{\mathcal{K}^{\prime}}\boxed{\underline{\psi(\mathcal{K}}\bm{\theta}_{\tau_k}[\underline{\mathcal{K}^{\prime}}}\underline{\psi(\mathcal{K}}(\bm{m}_{k}))])],\\ \bm{\omega}_{k,3}&=\underline{\mathcal{K}^{\prime}}\underline{\psi(\mathcal{K}}[\boxed{\underline{\mathcal{K}^{\prime}}\bm{\theta}_{\tau_k}\underline{\psi(\mathcal{K}}}[\underline{\mathcal{K}^{\prime}}\underline{\psi(\mathcal{K}}(\bm{m}_{k}))])].
    \end{aligned}$$

    It is obvious that the above analysis can easily be generalized to any $n$. And we need to define the left- and right-hand operators for odd and even numbers to distinguish between the two designs.

\section{Two concrete operator spectral geometric decomposition examples} \label{appendix-B}

Especially if $\phi_\ell(r)=\frac{1}{2}\|r\|^{2}$, we can obtain the first-order derivative $\phi^\prime_\ell\left(\mathcal{K}(\boldsymbol{x})\right)=\frac{d\phi_\ell\left(\mathcal{K}(\boldsymbol{x})\right)}{d\bm{x}}=\mathcal{K}^{\prime} \left( \mathcal{K} (\bm{x})\right)$ of the regularization term $\mathcal{R}(\bm{x})$ on $\mathbb{R}^{N}$, hence one has
\begin{equation}
    \mathcal{M}^i(\bm{m}_{k})=\underbrace{\underline{\mathcal{K}^{\prime}\mathcal{K}}[\cdots\underline{\mathcal{K}^{\prime}\mathcal{K}}[\underline{\mathcal{K}^{\prime}\mathcal{K}}}_{i}(\bm{m}_{k}]],
    \label{eq7-1}
\end{equation}
for $s=1$ and $i=1,\dots,n$.

Alternatively if $\phi_\ell(r)=\|r\|$, thus the first-order derivative of $\phi_\ell\left(\mathcal{K}(\boldsymbol{x})\right)$ in the regularization term $\mathcal{R}(\bm{x})$ defined on $\mathbb{R}^{N}$ is derived as follow
\[\phi^\prime_\ell\left(\mathcal{K}(\boldsymbol{x})\right)=\frac{d\phi_\ell\left(\mathcal{K}(\boldsymbol{x})\right)}{d\bm{x}}=\mathcal{K}^{\prime} \left( \frac{\mathcal{K} (\bm{x})}{|\mathcal{K} (\bm{x})|}\right),\] we define $\mathcal{M}^i(\bm{m}_{k})$ for $s=1$ by
\begin{equation*}
    \mathcal{M}^i(\bm{m}_{k})=
    \underbrace{\underline{\mathcal{K}^{\prime}
            \frac{\mathcal{K}}{\|\mathcal{K} (\bm{r}_{k,i})\|}
        }[\cdots\underline{\mathcal{K}^{\prime}
            \frac{\mathcal{K}}{\|\mathcal{K} (\bm{r}_{k,1})\|}
        }[\underline{\mathcal{K}^{\prime}
            \frac{\mathcal{K}}{\|\mathcal{K} (\bm{m}_k)\|}
    }}_{i}(\bm{m}_{k})]],
    \label{eq7-1-1}
\end{equation*}
where $i=1,\dots,n$, and the normalizing weight technique is adopted
to prevent the scale of features from becoming too large. Inspired by it, we can restrict the input of features by the above architecture and realize channel normalization as in the BN layer.

\bibliographystyle{IEEEtran}
\bibliography{Ref_Nest_DGIL}

\end{document}